\documentclass[preprint,12pt,authoryear]{elsarticle}

\usepackage[utf8]{inputenc}
\usepackage{amssymb}
\usepackage{amsmath}
\usepackage{booktabs}
\usepackage[normalem]{ulem}
\usepackage{theorem}
\usepackage{bbm}
\usepackage[pdftex, colorlinks]{hyperref}
\hypersetup{
    colorlinks = true,
    citecolor = {blue},
}
\numberwithin{equation}{section} 
\newtheorem{teo}{Theorem} 
\newtheorem{definition}{Definition} 
\journal{Neurocomputing}

\begin{document}

\begin{frontmatter}

\title{Consistent model selection for estimating functional interactions among stochastic neurons with variable-length memory}

\author[UFSCar]{Ricardo F. Ferreira}
\author[UFSCar]{Matheus E. Pacola}
\author[UFSCar]{Vitor G. Schiavone}
\author[FAU1]{Rodrigo F. O. Pena}

\address[UFSCar]{Department of Statistics, Federal University of S\~{a}o Carlos, S\~{a}o Carlos, SP, Brazil, 13565-905}
\address[FAU1]{Department of Biological Sciences, Florida Atlantic University, Jupiter, FL, USA, 33458}

\begin{abstract}
We address the problem of identifying functional interactions among stochastic neurons with variable-length memory from their spiking activity, \textcolor[rgb]{0,0,0}{where ``variable-length memory'' implies that the influence of past spikes can extend over time periods whose length itself may change, reflecting adaptive or context-dependent history effects.} The neuronal network is modeled by a stochastic system of interacting point processes with variable-length memory, \textcolor[rgb]{0,0,0}{meaning that each neuron's firing probability depends on its own and other neurons' historical spikes, with the length of this history not being fixed.} {Each chain describes the activity of a single neuron, indicating whether it spikes at a given time. One neuron's influence on another can be either excitatory or inhibitory. To identify the existence and nature of an interaction between a neuron and its postsynaptic counterpart,} we propose a model selection procedure based on the observation of the spike activity of a finite set of neurons over a finite time. The proposed procedure is also based on the maximum likelihood estimator for the synaptic weight matrix of the network neuronal model. In this sense, we prove the consistency of the maximum likelihood estimator {followed} by a proof of the consistency of the neighborhood interaction estimation procedure \textcolor[rgb]{0,0,0}{ensuring that, with enough data, the method accurately recovers both the values of the synaptic weights and the presence or absence of connections}. The effectiveness of the proposed model selection procedure is demonstrated using simulated data, which validates the underlying theory \textcolor[rgb]{0,0,0}{showing that, under controlled conditions, the estimated connections match the true simulated network, thereby confirming the accuracy and robustness of the approach}. The method is also applied to analyze spike train data recorded from hippocampal neurons in rats during a visual attention task, where a computational model reconstructs the spiking activity and the results reveal interesting and biologically relevant information.
\end{abstract}


\end{frontmatter}



\section{Introduction}
\label{sec1}

One of the most important concerns in modern neuroscience is to understand how animal behavior emerges from interactions between neurons and ensembles of neurons \citep{dayan2005theoretical,gerstner2014neuronal}. Until recently, the dominant paradigm in neuroscience was to measure the activity of a single neuron or a brain area to correlate it with animal behavior \citep{nicolelis2006seeking}. Advances in multichannel and optical technologies now enable researchers to record the activity of thousands of neurons simultaneously over several days \citep{brown2004multiple, nicolelis2006seeking, li2010scale, takahashi2010circuit, grewe2010high, ahrens2013whole, prevedel2014simultaneous}. \textcolor[rgb]{0,0,0}{In addition to spike-based analyses, extensive research has been conducted on the functional connectivity in EEG, such as through supervised network-based fuzzy learning \citep{yu2019supervised}, and the modulation of spectral power and functional connectivity \citep{yu2018modulation, yu2019modulation}. Functional magnetic resonance imaging (fMRI) allows the} recording global brain activity over extended hours \citep{logothetis2007ins}. Consequently, the challenge lies in using these data sets to understand the interactions among neurons and how these relate to animal behavior \citep{brown2004multiple, schneidman2006weak}. To address this challenge, we need methods that capture variability in neural activity, \textcolor[rgb]{0,0,0}{make accurate predictions, and provide interpretable representations of large-scale neural data.}

Experiments demonstrate that, generally, a specific animal behavior does not correspond to a unique pattern of neuronal activity. In fact, recordings of electrophysiological patterns both {\it in vitro} and {\it in vivo} reveal that neuronal activity is spontaneous, highly irregular \citep{stein2005neuronal, crochet2011synaptic, naud2012performance}, and variable in its response to certain stimuli \citep{bair1996temporal, nawrot2008measurement}. These observations indicate that neurons, synapses, and the entire neural system inherently exhibit stochastic properties \citep{schneidman1998ion, oram1999stochastic, buesing2011neural}. Consequently, the description of neural phenomena requires a probabilistic framework. Within this stochastic framework, each type of animal behavior corresponds to a specific probability distribution defined on the set of possible neural activity realizations. These realizations are characterized not only by the ensemble of spike trains but also by the time-evolving functional interactions. In this sense, the evolution of neuronal activity over time can be modeled as a countable system of interacting stochastic processes.

The stochastic approach has the advantage of incorporating part of the available knowledge about neural systems to construct parsimonious probabilistic models. However, determining which stochastic processes are more suitable for modeling neuronal activities in a network remains a matter of debate. In recent years, many probabilistic models have been proposed \citep{deco2009stochastic, harrison2005stochastic, toyoizumi2009mean, cessac2011statistics, stevenson2011advances, sacerdote2013stochastic, cofre2014exact, chevallier2017mean}, generally involving either Gibbsian or full-memory Markovian descriptions {(but see \citealp{lima2021modeling} for a more simplified stochastic approach)}. Some works in the literature have shown that these descriptions are inadequate \citep{friston2010free, truccolo2010collective, cessac2010view}. \textcolor[rgb]{0,0,0}{The fact that a neuron's membrane potential is reset to a resting level upon spiking implies that its time evolution depends on a variable-length history. More precisely, it is influenced by the input received from its presynaptic neurons since its last spike. Consequently, the system's dynamics cannot be described by a Markov process \citep{galves2016modeling}. In particular, under a continuous-time framework, the interspike intervals of a single neuron are not exponentially distributed, and the timing of each spike is influenced by the activity of neighboring neurons, which, in turn, depends on the collective configuration of the spike trains \citep{brillinger1988maximum}. This type of dependency does not align with a Markovian or Gibbsian description. Furthermore, experiments  with neurons suggest that their connections form sparse interaction graphs \citep{van2011rich, ercsey2013predictive} that, at least locally, differ from the graphs widely used in bioinformatics.} Therefore, the activity of a neuronal network could reasonably be modeled by large numbers of interacting point processes, with an interaction graph that varies over time and depends in a variable manner on the system's history \citep{galves2013infinite, galves2016modeling}. 
\textcolor[rgb]{0,0,0}{This variable-length memory structure is also advantageous from an estimation perspective, as it leads to the aggregation of certain transition probabilities that would otherwise be treated separately in a fixed-order Markov model. These characteristics justify our choice to study a model with a variable-length memory structure in this work over alternative neuronal models.}

In this context, the stochastic neuronal network that we consider is an example of a space-time model, called \textit{interacting chains with memory of variable length}, which are natural generalizations of the chains with variable-length memory (see, e.g., \citealp{rissanen1983universal, buhlmann1999variable, galves2008stochastic}). In this network, at {a given} time step, each neuron spikes with a probability that is an increasing function of its membrane potential. The membrane potential of a {particular} neuron is affected by the actions of all other neurons interacting with it. More precisely, the membrane potential of a neuron depends on the accumulated influence received from its presynaptic neurons since its last spike time. Whenever a neuron fires, its membrane potential is reset to a resting level, and at the same time, postsynaptic current pulses are generated, modifying the membrane potential of all its postsynaptic neurons. The contribution of a presynaptic neuron to the membrane potential of a postsynaptic neuron is either excitatory or inhibitory, depending on the sign of the synaptic weight from the pre- to the postsynaptic neuron. This description leads to a parsimonious understanding of the fundamental mechanisms underlying the functions of the nervous system at different scales. 
 
In the literature, numerous studies have focused on the probabilistic analysis of these models. Initially developed by \cite{galves2013infinite}, the GL neuron model is a discrete-time version of the integrate-and-fire (IF) model, featuring random thresholds and exponential-type postsynaptic current pulses. This situates it within a classical and widely accepted framework in modern neuroscience, supported by seminal works such as those by \cite{hodgkin1952quantitative, dayan2005theoretical, gerstner2002spiking, adrien1928basis, adrian1929discharge, gerstner1992associative, gerstner1995time}. As examples, we also may cite \cite{okatan2005analyzing, reynaud2014goodness, truccolo2005point, cofre2014exact} for works in a discrete-time framework. Additionally, there are approaches, similar to the current one, where continuous time is used \citep{de2015hydrodynamic, duarte2015hydrodynamic, duarte2016model, fournier2016toy, robert2016dynamics, yaginuma2016stochastic, chevallier2017mean, hodara2017hawkes}. 

In neuronal networks, {interactions between neurons are determined by how a neuron is connected to its neighbors, which consist of all their pre- and postsynaptic neurons}. A key challenge in such networks is estimating these interaction neighborhoods. Although neural activity can be directly observed, the interactions between neurons must be inferred from data. Traditionally, this has been done using descriptive statistical methods, such as linear correlation, which provide limited insights into the mechanisms driving neural activity dynamics \citep{bryant1973correlations, knox1974cross, brown2004multiple, platkiewicz2021monosynaptic}. Alternative approaches include the use of models developed in statistical mechanics, such as the Ising model, to infer neural interactions \citep{schneidman2006weak, galves2015identifying, lerasle2016sharp}. However, interpreting these results can be challenging because the Ising model does not closely resemble known {biophysical neuronal processes, although it was instrumental in the development of artificial neural networks such as the Hopfield model \citep{hopfield1982neural}}. Therefore, the main goal of this article is to introduce a straightforward statistical selection procedure for the aforementioned class of stochastic models, aimed at estimating the interaction neighborhood.

{The primary objective of identifying the underlying network structure from a neuronal system based on observed data is to discern its potential functional role.} These connections, whether dependencies, correlations, or causal relationships among neuronal entities, can be represented as directed edges in a graph, with associated synaptic weights delineating the strength of dependency from pre- to postsynaptic neurons. Many algorithms have been proposed to estimate both the network structure and the edge weights from time series data generated by various dynamic processes. Classical model selection methods for discrete graphical models or Markov random fields on graphs have been, for example, advanced by \cite{lauritzen1996graphical, csiszar2004consistent, koller2009probabilistic, pensar2017structure, divino2000penalized}. More recently, significant efforts have been directed towards estimating interaction graphs underlying models like finite volume Ising models \citep{montanari2009graphical, ravikumar2010high, bresler2013reconstruction, bresler2015efficiently}, infinite volume Ising models \citep{galves2015identifying, lerasle2016sharp, talata2014markov}, and variable-neighborhood random fields \citep{locherbach2011neighborhood}. From another perspective, graphical models can also be viewed as non-homogeneous versions of general random fields or Gibbs distributions on lattices \citep{georgii2011gibbs, comets1992consistency, comets1992parameter}. However, their application to stochastic modeling of neuronal data encounters a significant challenge: the assumption that the configuration describing neuronal activity at a given time conforms to a Gibbsian distribution lacks biological support \citep{cerqueira2017test}. As far as we know, this Gibbsian assumption lacks any biological grounding. Consequently, the methods typically employed for these graphical models only offer approximations of the true underlying distribution. 

The pursuit of statistical methods for estimating and selecting interaction graphs in dependency scenarios, such as those present in the neuronal model under consideration here, has its origins in neuroscience, likely beginning with \cite{brillinger1979empirical} and \cite{brillinger1988maximum}. Recent technological advancements now allow for the simultaneous recording of activity from thousands of neurons over extended periods. Consequently, statistical methods have been developed to accommodate fully \citep{pouzat2009automatic, ravikumar2010high} and partially observed networks \citep{lerasle2016sharp, duarte2015hydrodynamic, ost2020sparse, de2022estimating}. To our knowledge, distinct approaches have been taken to address the inference of interaction graphs for systems of neurons with variable-length memory, notably by \cite{duarte2015hydrodynamic, ost2020sparse, de2022estimating, izzi2024identifying}. However, despite the intriguing mathematical implications of the findings in \cite{duarte2015hydrodynamic}, the interaction neighborhood of a given neuron is estimated by assuming that we observe more neurons than this neighborhood even if it is not the totality of the network. In practice, the complexity of the algorithm makes it difficult to apply it to large data sets. In \cite{de2022estimating}, the authors overcome these drawbacks. \cite{ost2020sparse} propose a different approach, utilizing $\ell_1$-regularized regression to regress each variable on the remaining variables, and utilizing the sparsity pattern of the regression vector to infer the underlying neighborhood structure. Despite the mathematical significance of their findings, an experimental study was not conducted, thus hindering a comprehensive understanding of the method's efficacy in practice.

In this paper, we address the problem of estimating interaction neighborhoods based on the premise that neuronal activity is modeled using a space-time framework inspired by the Galves and L\"{o}cherbach model \citep{galves2013infinite}. This model is founded on the biologically plausible assumption that each neuron's membrane potential is reset every time it spikes. Leveraging well-established statistical principles, we first formulate and examine the consistency properties of maximum likelihood (ML) parameter estimation for this neuronal model. For each neuron $i$ in the sample, the proposed statistical selection procedure estimates the spiking probability vector based on the spike trains of all other neurons since its last spike time using the ML principle. For each neuron $j \ne i$, the estimated spiking probability vector is obtained without considering neuron $j$ in the sample. We then calculate a sensitivity measure between these estimated probability vectors. If this measure is statistically insignificant, we conclude that neuron $j$ does not belong to the interaction neighborhood of neuron $i$. A second contribution of this paper is a detailed analysis of the statistical consistency of this method. The effectiveness of the proposed model selection procedure is demonstrated using simulated data, {which validates} the underlying theory. \textcolor[rgb]{0,0,0}{The method is also applied to analyze spike train data recorded from hippocampal neurons in animals during a visual attention task, where a reconstruction using a simple network populated by leaky integrate-and-fire neuron models reveals interesting and biologically relevant information.}

The remainder of this article is organized as follows. In the next section, we highlight the experimental significance. In section \ref{sec3}, we establish our notations.  In Section \ref{sec4}, we review preliminary definitions and concepts, particularly those concerning the neuronal network model. In Section \ref{sec5}, we introduce the synaptic weight matrix estimation procedure and state our first result (Theorem \ref{thm:MLE_consistency}). In Section \ref{sec6}, we propose a new interaction neighborhood estimation procedure and state our second result (Theorem \ref{thm:model_selection_consistency}). In Section \ref{sec7}, we apply the proposed methodology to the identification of connectivity among stochastic neurons using synthetic data generated from the random network model described in Section \ref{sec4}. In Section \ref{sec8}, we apply the methodology to real data obtained from electrophysiology. The proof of Theorems \ref{thm:MLE_consistency} and \ref{thm:model_selection_consistency} are presented in Section \ref{sec9}. Lastly, we end this article with our conclusions in Section \ref{sec10}.

\section{Experimental significance}
\label{sec2}

Recent advancements in experimental techniques for recording and stimulating neuronal activity, including genetic manipulations, multi-electrode arrays, optogenetics, and voltage imaging, have significantly improved our access to a wide variety of neurons with increased precision. However, despite these advancements, the {\it in vivo} environment remains highly stochastic, complicating the reliable inference of functional connectivity without robust analytical approaches. In this study, we address these challenges by focusing on multichannel electrophysiological recordings from the CA1 region in rats. From these recordings, we selected five neurons, interneurons or pyramidal cells, without prior knowledge of their connectivity. Our approach enables the estimation of connectivity matrices, which we further analyze and simulate through a simplified computational model of these cells. Our analysis leverages a synergistic combination of experimental and theoretical methods, coupled with computational simulations, allowing each approach to inform and strengthen the other.

\section{Notations}
\label{sec3}

In this paper, we denote random variables in uppercase letters, stochastic chains in uppercase bold letters, and the specific values assumed by them in lowercase letters. Calligraphic letters denote the alphabets where random variables take values. Subscripts denote the outcome's position in a sequence; for example, $X_t$ generally indicates the $t^{th}$ outcome of the process $\boldsymbol{X}$. For any integers $j$ and $k$ such that $j \leq k$, we use the notation $x_j^k$ for finite sequences $\left(x_{j}, \ldots, x_k\right)$, $x_{-\infty}^k$ for left-infinite sequences $\left(\ldots, x_{k-1}, x_k\right)$, and $x_{k}^{+\infty}$ for right-infinite sequences $\left(x_k, x_{k+1}, \ldots\right)$. We use the convention that if $j > k$, $x_j^k$ is the empty sequence. We use analogous notations for sequences of random variables. 

The cardinality of a set $\mathcal{V}$ is denoted by $|\mathcal{V}|$. We write $\mathbb{N}$ to denote the set of natural numbers $\{0, 1, 2, \ldots\}$, $\mathbb{Z}$ to denote the set of integer numbers $\{\ldots, -1, 0, 1, \ldots\}$, $\mathbb{Z}_{-}$ for the set of negative integers and $\mathbb{Z}_{+}$ for the set of positive integers. For $m, n \in \mathbb{N}$, we denote by $M_{m \times n}(\mathbb{R})$ the set of all $m \times n$ matrices with real entries. Finally, $\mathbb{I}\{\cdot\}$ stands for the indicator of a set or event.

\section{Neuronal Network Model}
\label{sec4}

Neurons are electrical cells communicating among themselves via the emission of action potentials, also called \textit{spikes}. The sequence of times at which an individual neuron in the nervous system generates an action potential is termed a \emph{spike train}. We adopt here a discrete time approach to model spike train data. In this approach, it is useful to consider the times of spike occurrence with a certain degree of accuracy, which is called the \emph{{bin size}} \citep{mackay1952limiting}. {In other words, the bin size} refers to the duration of time over which neural activity is aggregated or binned for analysis.  For a small enough bin size ($10$ ms is a typical choice), the spike train may be represented as a binary sequence $x_{1}^{n} \in \{0,1\}^n$, where
$$x_t = \begin{cases}
1, &\hbox{ if the neuron spikes at the } t^{th} \hbox{ bin}, \\
0, &\hbox{ otherwise},
\end{cases}$$
for all $t = 1, 2, \ldots, n$. The appropriate bin size to use depends on the specific experimental design and the characteristics of the data being analyzed. In general, the bin size is chosen to {allow for a balance between capturing relevant} details of the neuronal activity and having sufficient statistical power. This typically involves selecting a bin size that is small enough to capture important features of the data but not so small that the resulting spike counts are noisy or unreliable.

Recordings of neuronal activity reveal irregular spontaneous {firing of neurons and variability in their response to the same stimulus \citep{hill1929basis, adrian1929discharge, gerstner1992associative, gerstner1995time,Lin09}, also known as trial-to-trial variability}. Thus, the experimental data suggest that spike trains should be modeled from a probabilistic point of view. In this context, and to give a probability measure to describe the process of spiking as a sequential process, we assume that the network is represented by a discrete-time homogeneous stochastic chain $\boldsymbol{X} := \left\{X_t: t \in \mathbb{Z}\right\}$ defined on a suitable probability space $(\Omega, \mathcal{F}, \mathbb{P})$, where
$$X_t =  \begin{cases} 
1, & \hbox{ if the neuron spikes at the }  t^{th} \hbox{ bin},\\
0, & \hbox{ otherwise},
\end{cases}$$
for every $t \in \mathbb{Z}$. 

In this paper, we assume that the sample spike train is generated by a stochastic source. This means that at each bin, conditional on the whole past, there is a fixed probability of obtaining a spike. Neurons exhibiting this characteristic are arranged in such a way that they share similar biophysical properties and are collectively referred to as \emph{{stochastic neurons}}.

Let $I$ be a finite set of stochastic neurons, and assume that the bins are indexed by the set $\mathbb{Z}$.  In this context,  the network of neurons is described by a discrete-time homogeneous stochastic chain $\boldsymbol{X} := \left\{X_t(i) : i \in I, t \in \mathbb{Z}\right\}$. For each neuron $i \in I$ at each bin $t \in \mathbb{Z}$, 
$$X_t(i) = \begin{cases}
1, &\hbox{ if  neuron } i \hbox{ spikes at the }  t^{th} \hbox{ bin},\\
0, &\hbox{ otherwise}.
\end{cases}$$
Moreover, whenever we say time $t \in \mathbb{Z}$, it should be interpreted as time bin $t$. For notational convenience, we write $\boldsymbol{X}_t(F) = \left\{X_t(i): i \in F\right\}$ to represent the configuration of $\boldsymbol{X}$ at time $t \in \mathbb{Z}$, restricted to the neuron set $F \subset I$, and the path of $\boldsymbol{X}$ from $t-\ell$ to $t-1$, restricted to the neuron set $F \subset I$, as $\boldsymbol{X}_{j}^{k}(F) = \left\{X_{t}(i): i \in F, j \leq t \leq k\right\}$, where $j$ and $k$ are positive integers such that $j \leq k$. When $F = I$, we will simply write $\boldsymbol{X}_t$ and $\boldsymbol{X}_{j}^{k}$, respectively. We use analogous notation for the observed configuration of $\boldsymbol{X}$ and the observed path of $\boldsymbol{X}$. 

In the network with stochastic neurons considered in this article, the stochastic chain $\boldsymbol{X}$ has the following dynamic. At each time step, conditional on the whole past, neurons update independently from each other, i.e., for any $t \in \mathbb{Z}$, any $F \subset I$ and any choice $x_t(i) \in \{0,1\}$, $i \in F$, we have $\mathbb{P}$-a.s.,
\begin{equation}
\mathbb{P}\left(\left.\bigcap_{i \in F} \left\{{X}_t(i) = {x}_t(i)\right\}\right|\boldsymbol{X}_{-\infty}^{t-1} = \boldsymbol{x}_{-\infty}^{t-1}\right) = \prod_{i \in F} \mathbb{P} \left(\left.X_t(i) = x_t(i)\right|\boldsymbol{X}_{-\infty}^{t-1} = \boldsymbol{x}_{-\infty}^{t-1}\right),
\label{eq:independence}
\end{equation}
where $\boldsymbol{x}_{-\infty}^{t-1}$ is a left-infinite configuration of $\boldsymbol{X}$.

Moreover, the probability that neuron $i \in I$ spikes at bin $t \in \mathbb{Z}$, conditional on the whole past, is a non-decreasing measurable function of its membrane potential. In other words, for each neuron $i \in I$ at any $t \in \mathbb{Z}$,
\begin{equation}
\mathbb{P}\left(\left.X_{t}(i) = 1\right| \boldsymbol{X}_{-\infty}^{t-1} = \boldsymbol{x}_{-\infty}^{t-1}\right) = \phi_i\left(v_{t-1}(i)\right)
\label{eq:transition_probability1}
\end{equation}
$\mathbb{P}$-a.s., where $v_t(i) \in \mathbb{R}$ denotes the membrane potential of neuron $i \in I$ at time $t \in \mathbb{Z}$ and $\phi_i: \mathbb{R} \rightarrow [0,1]$ is a non-decreasing function called the \emph{spiking rate function}. 

The membrane potential of a given neuron $i \in I$ is affected by the actions of all other neurons interacting with it. More precisely, the membrane potential of a given neuron $i \in I$ depends on the influence received from its presynaptic neurons since its last spiking time. In this sense, the probability of neuron $i \in I$ spiking increases monotonically with its membrane potential. Whenever neuron $i \in I$ fires, its membrane potential is reset to a resting value, and at the same time, postsynaptic current pulses are generated, modifying the membrane potential of all its postsynaptic neurons. When a presynaptic neuron $j \in I - \{i\}$ fires, the membrane potential of neuron $i \in I$ changes. The contribution of neuron $j \in I$ to the membrane potential of neuron $i \in I$ is either excitatory or inhibitory, depending on the sign of the synaptic weight of neuron $j$ on neuron $i$. Moreover, the membrane potential of each neuron in the network is affected by the presence of leakage channels in its membrane, which tends to push its membrane potential toward the resting potential. This spontaneous activity of neurons is observed in biological neuronal networks.

Assuming the above description, we may consider stochastic neurons with several kinds of short-term memory.  In this article, we explore a stochastic neuronal model based on the discrete version of the GL model \citep{galves2013infinite}, where neuronal spike trains are prescribed by interacting chains with variable-length memory. In this model, for each neuron $i \in I$ at any bin $t \in \mathbb{Z}$, we can write
$$v_{t-1}(i) = \begin{cases}
0, &\hbox{ if } x_{t-1}(i) = 1, \\
\sum_{j \in I} \omega_{j\rightarrow i} \sum_{s = L_{t}^{(i)} + 1}^{t-1} \dfrac{x_{s}(j)}{2^{t - L_t^{(i)}-1}} , &\hbox{ otherwise},
\end{cases}$$
where $\omega_{j \rightarrow i} \in \mathbb{R}$ is the synaptic weight of neuron $j$ on neuron $i$ and $L_t^{(i)}$ is the last spike time of neuron $i \in I$ before time $t \in \mathbb{Z}$, i.e.,
$$L_t^{(i)} := \sup\left\{s < t : x_s(i) = 1\right\}, \quad \forall i \in I.$$
{Therefore}, for each neuron $i \in I$ at any $t \in \mathbb{Z}$, we may rewrite \eqref{eq:transition_probability1}, $\mathbb{P}$-a.s., in the following way
\begin{equation}
\mathbb{P}\left(\left.X_{t}(i) = 1\right| \boldsymbol{X}_{-\infty}^{t-1} = \boldsymbol{x}_{-\infty}^{t-1}\right) = \phi_i\left(\left(1 - x_{t-1}(i)\right)\left(\sum_{j \in I} \omega_{j \rightarrow i} \sum_{s = L_{t}^{(i)} + 1}^{t-1} \dfrac{x_{s}(j)}{2^{t - L_t^{(i)}-1}}\right)\right).
\label{eq:transition_probability3}
\end{equation}

Since the spike rate function $\phi_i$ is non-decreasing for any $i \in I$, spikes from a presynaptic neuron $j \in I - \{i\}$ excite neuron $i$ when $w_{j \rightarrow i} > 0$. In contrast, if $w_{j \rightarrow i} < 0$, spikes from the presynaptic neuron $j$ inhibit neuron $i$. Finally, if $w_{j \rightarrow i} = 0$, neuron $j$ has no influence on neuron $i$, that is, $j$ does not belong to the interaction neighborhood of neuron $i$.  We suppose that $w_{i \rightarrow i} = 0$ for any $i \in I$. Note that since $I$ is a finite set of neurons, any family of synaptic weights has the following property of uniform summability:
\begin{equation}
r := \sup_{i \in I} \sum_{j \in I} |w_{j \rightarrow i}| < \infty.
\label{assumption1:uniform_summability}
\end{equation}

In this version of GL neuronal model, we define the \textit{leak functions} $g_i: (0, +\infty) \rightarrow [0, +\infty)$ in the following way
\begin{equation}
g_i(t-s) := \dfrac{1}{2^{t - L_t^{(i)} - 1}},
\label{assumption2:leak_function_finite}
\end{equation}
for all $s = L_{t}^{(i)}+ 1, \ldots, t - 1$, $t \in \mathbb{Z}$ and $i \in I$. These functions describe how neuron $i$ loses potential due to leakage effects over time. In fact, note that if a presynaptic neuron $j \in I - \{i\}$ fires a fixed number of times from the last spike of the postsynaptic neuron $i \in I$ until time $t-1$, then the contribution of neuron $j$ on the membrane potential of neuron $i$ is smaller the further back the last spike of the postsynaptic neuron occurred. Therefore, the presence of leakage channels tends to push the postsynaptic membrane potential toward the resting potential.

Observe that the spiking probability of a given neuron depends on the accumulated activity of the system after its last spike time. Here, we adopt the convention that $L_t^{(i)} \geq t - K$, where $K$ is a positive integer number that represents the largest memory length of all stochastic neurons considered in the network. This implies that the time evolution of each single neuron looks like a Markov chain with variable-length memory. In this sense, the model considered in this article is slightly different from the GL neuronal model, in which the memory of the process can depend on an infinite past. Thus, here $L_t^{(i)} = t - K$, when $x_s(i) = 0$ for every $s < t$. This structure of variable-length memory is more appropriate from the estimation point of view because it implies that some transition probabilities of the Markov chain with order $K$ are lumped together.

Assuming that for all $i\in I$, the spike rate function $\phi_i$ is strictly increasing and uniformly Lipschitz continuous such that there exists a real number $\delta \in ]0, 1[$ 
\begin{equation}
\delta \leq \phi_i(v) \leq 1 - \delta,
\label{assumption3:spike_rate_function_bounded}
\end{equation}
one can show the existence and uniqueness of a stationary stochastic chain $\boldsymbol{X}$ satisfying \eqref{eq:independence} whose dynamics are given by \eqref{eq:transition_probability3}. We refer the interested reader to \cite{galves2013infinite} for a rigorous proof of this result in the general version of the GL neuronal model. We also suggest reading \cite{ferreira2020non}, where results on the existence and uniqueness of stationary chains of this nature are also presented under more general assumptions.

\section{Synaptic weights estimation}
\label{sec5}

The interaction neighborhood estimation procedure presented in this article is based on the maximum likelihood (ML) estimator of the synaptic weight matrix
$$W := \begin{bmatrix}
\omega_{1 \rightarrow 1} & \omega_{1 \rightarrow 2} & \ldots & \omega_{1 \rightarrow N} \\
\omega_{2 \rightarrow 1} & \omega_{2 \rightarrow 2} & \ldots & \omega_{2 \rightarrow N} \\
\vdots & \vdots & \ddots & \vdots \\
\omega_{N \rightarrow 1} & \omega_{N \rightarrow 2} & \ldots & \omega_{N \rightarrow N}
\end{bmatrix} \in M_{N \times N}(\mathbb{R}).$$
defined by stochastic neurons with variable-length memory that follow the dynamic of the model described in the previous section. In this section, we recover the neuronal network model parameters via ML estimation and establish the strong consistency of the ML estimators.

Consider a finite network of neurons and a sample of spike trains over a finite time horizon. Given positive integers $N$ and $T$, such that $N < T$, define $|I| = N$ as the number of neurons sampled over $T$ bins. Thus, given the observed configuration $\boldsymbol{x}_{-K+1}^{T} \in \{0,1\}^{N \times (T+K)}$ of $\boldsymbol{X}$, the rescaled log-likelihood function is defined as 
\begin{align*}
\ell\left(W; \boldsymbol{x}_{-K+1}^T\right) &:= \dfrac{1}{T} \sum_{t = 1}^{T} \log \mathbb{P}\left\{\left. \boldsymbol{X}_t = \boldsymbol{x}_t\right| \boldsymbol{X}_{-K+1}^{t-1} = \boldsymbol{x}_{-K+1}^{t-1}\right\} \\
&+ \dfrac{1}{T} \log \mathbb{P}\left\{\boldsymbol{X}_{-K+1}^{0} = \boldsymbol{x}_{-K+1}^{0}\right\},
\end{align*}
where the rescaling factor $1/T$ in this definition is for later theoretical convenience. Now it turns out that the term $\dfrac{1}{T} \log \mathbb{P}\left\{\boldsymbol{X}_{-K+1}^{0} = \boldsymbol{x}_{-K+1}^{0}\right\}$ is dominated by the others as $T$ goes to infinity. If we assume that for each $W$ the initial distribution $\mathbb{P}\left\{\boldsymbol{X}_{-K+1}^{0} = \boldsymbol{x}_{-K+1}^{0}\right\}$ is independent of the model parameters, then $\ell(W; \boldsymbol{x}_{-K+1}^T)$ can be rewritten as
$$\ell(W; \boldsymbol{x}_{-K+1}^T) =  \dfrac{1}{T} \sum_{t = 1}^{T} \log \mathbb{P}\left\{\left. \boldsymbol{X}_t = \boldsymbol{x}_t\right| \boldsymbol{X}_{-K+1}^{t-1} = \boldsymbol{x}_{-K+1}^{t-1}\right\}.$$
In particular, we can assume that the initial distribution has all its mass concentrated at $\boldsymbol{x}_{-K+1}^0$. For this reason, the term containing the initial distribution can be omitted. Such functions will still be called log-likelihood functions, but it is important to observe that in the literature these functions can also be found under the name pseudo-likelihood functions. 

Assuming that the firing rate function $\phi_i$ of the postsynaptic neuron $i \in I$ is the logistic function $\phi_i(v) = \frac{e^v}{1 + e^v}$ for all $v \in \mathbb{R}$, and following some algebraic manipulation, the rescaled log-likelihood function can be written as

\begin{align*}
\ell\left(W; \boldsymbol{x}_{-K+1}^T\right)  &= \dfrac{1}{T} \sum_{i = 1}^N\sum_{t=1}^T \left[x_t(i)\log\left(\frac{\phi_{i}\left(v_{t-1}(i)\right)}{1-\phi_{i} \biggl(v_{t-1}(i)\biggl)}\right)+\log\left(1-\phi_{i}\left(v_{t-1}(i)\right)\right)\right]\\
&= \dfrac{1}{T} \sum_{i = 1}^N \sum_{t=1}^{T}\biggl[x_t(i)v_{t-1}(i)-\log\left(1+\exp\left(v_{t-1}(i)\right)\right)\biggl] \\
&= \dfrac{1}{T} \sum_{i=1}^N \sum_{t=1}^{T} \left[ x_t(i)\left(1 - x_{t-1}(i)\right)\left(\sum_{j \in I} \omega_{j \rightarrow i} \sum_{s = L_{t}^{(i)} + 1}^{t-1} \dfrac{x_{s}(j)}{2^{t - L_t^{(i)}-1}}\right) \right. \\
&\left.- \log\left(1 + \exp\left(\left(1 - x_{t-1}(i)\right)\left(\sum_{j \in I} \omega_{j \rightarrow i} \sum_{s = L_{t}^{(i)} + 1}^{t-1} \dfrac{x_{s}(j)}{2^{t - L_t^{(i)}-1}}\right)\right)\right)\right] \\
&:= \sum_{i=1}^N \ell^{(i)}(\boldsymbol{\omega}^{(i)}, \boldsymbol{x}_{-K+1}^T),
\end{align*}
where $\ell^{(i)}(\boldsymbol{\omega}^{(i)}, \boldsymbol{x}_{-K+1}^T)$ are rescaled log-likelihood functions for the parameter vector $\boldsymbol{\omega}^{(i)} = \left[w_{1 \rightarrow i} \quad  w_{2 \rightarrow i} \quad \cdots \quad w_{N \rightarrow i}\right]^{\top}$ associated with the response of the postsynaptic neuron $i$ to the neighboring values.

The separability of the likelihood function allows us to estimate the coefficients associated with each postsynaptic neuron $i$
independently of the others. It is important to note that each independent log-likelihood utilizes all the available data. This separability simplifies the analysis and enables the use of embarrassingly parallel algorithms to estimate all the parameters. Essentially, this estimation procedure can be viewed as solving $N$ logistic regression problems.

With this set-up, for each postsynaptic neuron $i$, denote by $\mathcal{T}^{(i)}_T$ the set of all sequences $\boldsymbol{u}$ that appear at least once in the sample $\boldsymbol{x}_{-K+1}^{T}$, that is
$$\mathcal{T}^{(i)}_T := \left\{\boldsymbol{u} \in \bigcup_{\ell = 2}^{T + K}\{0,1\}^{N \times \{-\ell+1, -\ell+2, \ldots, -1\}}: N^{(i)}_T(\boldsymbol{u}) \geq 1\right\},$$
where $N^{(i)}_T(\boldsymbol{u})$ counts the number of occurrences of $\boldsymbol{u}$ in the sample $\boldsymbol{x}_{-K+1}^{T}$, when the last spike of neuron $i$ has occurred $\ell$ time steps before in the past, i.e., 
$$N_T^{(i)}(\boldsymbol{u}) := \sum_{t = \ell-K+1}^{T+1} \mathbb{I}\left\{\boldsymbol{X}_{t-\ell}^{t-1}(i) = 10^{\ell-1}, \boldsymbol{X}_{t-\ell+1}^{t-1}(I - \{i\}) = \boldsymbol{u}\right\}.$$

In this sense, we can write $\ell^{(i)}(\boldsymbol{w}^{(i)}, \boldsymbol{x}_{-K+1}^T)$ as
$$\ell^{(i)}(\boldsymbol{\omega}^{(i)}, \boldsymbol{x}_{-K+1}^T) = \sum_{\boldsymbol{u} \in \mathcal{T}_T^{(i)}}  \sum_{a = 0}^{1} \dfrac{N_T^{(i)}(\boldsymbol{u}, a)}{T} \log  P_{\boldsymbol{u}a}(\boldsymbol{\omega}^{(i)}),$$
where 
$$N_T^{(i)}(\boldsymbol{u}, a) :=  \sum_{t = \ell-K+1}^{T+1} \mathbb{I}\left\{\boldsymbol{X}_{t-\ell}^{t-1}(i) = 10^{\ell-1}, \boldsymbol{X}_{t-\ell+1}^{t-1}(I - \{i\}) = \boldsymbol{u}, X_t(i) = a\right\}$$ 
counts the number of occurences of $\boldsymbol{u}$ followed or not by a spike of neuron $i$ ($a = 1$ or $a = 0$, respectively) in the sample $\boldsymbol{x}_{-K+1}^{T}$, when the last spike of neuron $i$ has occurred $\ell$ time steps before in the past, and $P_{\boldsymbol{u}, a}(\boldsymbol{w}^{(i)})$ is the transition probability from $\boldsymbol{u} \in \mathcal{T}^{(i)}_T$ to $a \in \{0,1\}$ according to \eqref{eq:transition_probability3}.

We then define, for each postsynaptic neuron $i \in I$,  the ML estimator $\hat{\boldsymbol{w}}_T^{(i)}$ for the synaptic weight vector $\boldsymbol{w}^{(i)}$ by
\begin{equation}
\hat{\boldsymbol{\omega}}_T^{(i)} \in \arg\max_{\boldsymbol{\omega}^{(i)} \in \mathbb{R}^N} \ell^{(i)}(\boldsymbol{\omega}^{(i)}, \boldsymbol{x}_{-K+1}^T).
\label{eq:ML_estimator_definition}
\end{equation}
One of the results of this paper is the following consistency result for ML estimator $\hat{\boldsymbol{w}}_T^{(i)}$.

\begin{teo}
Let $I$ be a finite set of neurons, $K$ and $T$ be  positive integers, and $\boldsymbol{x}_{-K+1}^{T}$ be a sample produced by the stochastic chain $\boldsymbol{X}$ compatible with \eqref{eq:independence} and \eqref{eq:transition_probability1}, starting from $\boldsymbol{X}_{-K+1}^0 = \boldsymbol{x}_{-K+1}^{0}$ for some admissible past $\boldsymbol{x}_{-K+1}^{0} \in \Omega^{adm}$, where
$$\Omega^{adm} := \left\{\boldsymbol{x}_{-K+1}^0 \in \{0,1\}^{N \times K} : \forall i \in I, \exists \ell_i \hbox{ with } x_{\ell_i}(i) = 1\right\}.$$
Under assumptions \eqref{assumption1:uniform_summability}-\eqref{assumption3:spike_rate_function_bounded} , for any postsynaptic neuron $i \in F$, the following holds.
\begin{enumerate}
\item \textbf{(Strong consistency).} The ML estimator $\hat{\boldsymbol{\omega}}^{(i)}_T$ for $\boldsymbol{w}^{(i)}$ is strongly consistent, i.e., almost surely,
$$\lim_{T \rightarrow +\infty} \hat{\boldsymbol{\omega}}^{(i)}_T = \boldsymbol{w}^{(i)}_0,$$
where $\boldsymbol{w}^{(i)}_0  \in \mathbb{R}^N$ is the true parameter vector $\boldsymbol{w}^{(i)}$.
\item \textbf{($L^2$ consistency).} The ML estimator $\hat{\boldsymbol{\omega}}^{(i)}_T$ for $\boldsymbol{w}^{(i)}$ is $L^2$ consistent, i.e., 
$$\lim_{T \rightarrow +\infty}E\left[\left\|\hat{\boldsymbol{\omega}}^{(i)}_T - \boldsymbol{w}^{(i)}_0\right\|^2_2\right] = 0,$$
where $\boldsymbol{w}^{(i)}_0  \in \mathbb{R}^N$ is the true parameter vector $\boldsymbol{w}^{(i)}$ and $\|\cdot\|_2$ denotes the $\ell_2$-norm in $\mathbb{R}^N$.
\end{enumerate}
\label{thm:MLE_consistency}
\end{teo}

\section{Interaction neighborhood estimation procedure}
\label{sec6}

Consider the problem where we are interested in estimating the connectivity graph defined by stochastic neurons with variable-length memory that follow the dynamic of the model described in the Section \ref{sec2}. To do this, consider a finite network of neurons and a sample of spike trains over a finite time horizon. In this sense, given positive integers $N$ and $T$, define $|I| = N$ as the number of neurons sampled over $T$ bins. Thus, given the observed configuration $\boldsymbol{x}_{-K+1}^{T} \in \{0,1\}^{T+K}$ of $\boldsymbol{X}$,  we would like to estimate the interaction neighborhoods $\mathcal{V}_i$ of the sampled neurons $i \in I$, which is defined as 
$$\mathcal{V}^{(i)} = \{j \in I - \{i\}: \omega_{j \rightarrow i} \not= 0\},$$
i.e., the set of presynaptic neurons of $i$. 

The goal of our statistical selection procedure is to identify the set $\mathcal{V}_i$ from the data in a consistent way. Our procedure is based on the pseudo ML estimation of the synaptic weight matrix  $W \in M_{N \times N}(\mathbb{R}).$ Broadly speaking, in the statistical selection procedure we consider, we observe the network activity within a finite sampling region over a finite time interval. For each neuron $i$ in the sampled region, we estimate its spiking probability given the activity of all other neurons since the last spike of neuron $i$. For each neuron $j \not= i$, we introduce a sensitivity measure of this conditional firing probability with respect to the absence of presynaptic neuron $j$ in the network. If this sensitivity measure is statistically insignificant, we conclude that neuron $j$ does not belong to the interaction neighborhood $\mathcal{V}_i$ of neuron $i$. In the following, we rigorously define this statistical procedure.

For each postsynaptic neuron $i \in I$, we define the estimated probability of neuron $i$ spiking at time $t$, given the activity of neurons in the subset $F \subset I$ as
$$\hat{P}_t^{(i)}(F) := \phi_i\left((1 - x_{t-1}(i)) \left(\sum_{j \in F} \hat{\omega}_{j \rightarrow i} \sum_{s = L_t^{(i)} + 1}^{t-1} \dfrac{x_s(j)}{2^{2^{t - L_t^{(i)} - 1}}}\right)\right),$$
where $\hat{\omega}_{j \rightarrow i}$ is the pseudo maximum likelihood estimate of the synaptic weight from neuron $j$ on neuron $i$. This estimated probability is a predictor of neuron $i$ activity at time $t$ when influenced by the activity of all neurons in $F$. The vector $\hat{\boldsymbol{P}}^{(i)}_F := \left(\hat{P}_1^{(i)}(F), \hat{P}_2^{(i)}(F), \ldots, \hat{P}_T^{(i)}(F)\right)$ is, therefore, the vector of predicted firing probabilities over the $T$ time windows.

In particular, when $F = I - \{j\}$, i.e., the set of all neurons in $I$ except for neuron $j$, we can define a sensitivity measure $d: [0,1]^{T} \times [0,1]^{T} \rightarrow [0, +\infty[$ such that
 $$d\left(\hat{\boldsymbol{P}}_F^{(i)}, \hat{\boldsymbol{P}}_I^{(i)}\right) := \dfrac{1}{T}\sum_{t=1}^T \left|\hat{P}_t^{(i)}(F) - \hat{P}_t^{(i)}(I)\right|^2,$$
where $\|\cdot\|_{\ell^2}$ denotes the Euclidean norm. In this context, we say that neuron $j$ is a neighbor of neuron $i$ when $d(\hat{\boldsymbol{P}}_F^{(i)}, \hat{\boldsymbol{P}}_I^{(i)}) > \epsilon$  for some fixed cutoff point $\epsilon > 0$. Therefore, our interaction neighborhood estimator is defined as follows.

\begin{definition}
For any positive threshold parameter $\epsilon \in (0,1)$, the estimated interaction neighborhood of neuron $i \in I$, at accuracy $\epsilon$, given the sample $\boldsymbol{x}_{-K+1}^{T} \in \{0,1\}^{T+K}$ is defined as
$$\hat{\mathcal{V}}_{T}^{(i, \epsilon)} := \left\{j \in I - \{i\}: d(\hat{\boldsymbol{P}}_{I - \{j\}}^{(i)}, \hat{\boldsymbol{P}}_I^{(i)}) > \epsilon\right\}.$$
\end{definition}

The following theorem states the consistency of the interaction neighborhood estimator $\hat{\mathcal{V}}_{T}^{(i, \epsilon)}$ when $\mathcal{V}^{(i)} \subset I$.

\begin{teo}
Let $I$ be a finite set of neurons, $K$ and $T$ be positive integers, and $\boldsymbol{x}_{-K+1}^{T}$ be a sample produced by the stochastic chain $\boldsymbol{X}$ compatible with \eqref{eq:independence} and \eqref{eq:transition_probability1}, starting from $\boldsymbol{X}_{-K+1}^0 = \boldsymbol{x}_{-K+1}^{0}$ for some admissible past $\boldsymbol{x}_{-K+1}^{0} \in \Omega^{adm}$, where
$$\Omega^{adm} := \left\{\boldsymbol{x}_{-K+1}^0 \in \{0,1\}^{N \times K} : \forall i \in I, \exists \ell_i \hbox{ with } x_{\ell_i}(i) = 1\right\}.$$
Under assumptions \eqref{assumption1:uniform_summability}-\eqref{assumption3:spike_rate_function_bounded} , for any postsynaptic neuron $i \in F$, the following holds.
\begin{enumerate}
\item \textbf{(Overestimation).} For any $j \in I - \mathcal{V}^{(i)}$, we have that for any $\epsilon > 0$,
$$ \lim_{T \rightarrow +\infty} \mathbb{P}\left(j \in \hat{\mathcal{V}}^{(i, \epsilon)}_T\right) = 0.$$
\item \textbf{(Underestimation).} For any $j \in \mathcal{V}^{(i)}$, we have that for any $0 < \epsilon < m_i$,
$$ \lim_{T \rightarrow +\infty} \mathbb{P}\left(j \not\in \hat{\mathcal{V}}^{(i, \epsilon)}_T\right) = 0,$$
where 
$$m_i := \inf_{u \in D_i} \{\phi^{\prime}(u)\} \inf_{j \in \mathcal{V}^{(i)}}|\omega_{j \rightarrow i}|,$$
and
$$D_i := \left[\sum_{k \in \mathcal{V}_{-}^{(i)}} \omega_{k \rightarrow i}, \ \sum_{k \in \mathcal{V}_{+}^{(i)}} \omega_{k \rightarrow i}\right],$$
with $\mathcal{V}_{-}^{(i)} := \{k \in I: \omega_{k \rightarrow i} < 0\}$ and $\mathcal{V}_{+}^{(i)} := \{k \in I: \omega_{k \rightarrow i} > 0\}$.
\item \textbf{(Consistency).} For any $\epsilon > 0$, we have
$$\lim_{T \rightarrow +\infty} \mathbb{P}\left(\hat{\mathcal{V}}^{(i, \epsilon)}_T \not= \mathcal{V}^{(i)}\right) = 0.$$
\end{enumerate}
\label{thm:model_selection_consistency}
\end{teo}

\section{Results on simulation}
\label{sec7}

This section presents numerical experiments to verify the consistency of the ML estimator for the synaptic weight matrix and the interaction neighborhood estimation procedure. These experiments complement the theoretical results, highlighting the practical applicability of the underlying theory. The simulations were conducted in R, version 8.16, and the code, along with documentation, has been made publicly available on the author's GitHub for future use by the scientific community.

\subsection{Simulation setup}
\label{subsec61}

We conducted the simulation study considering four distinct scenarios. Each scenario is based on different neurobiological principles {such as the ratio of excitatory to inhibitory neurons and their synaptic weights, attempting to account for the complexity and variability as well as a balanced state inherent in neuronal behavior.} In this way, we were able to evaluate the robustness of the conclusions obtained, reducing the risk of oversimplifications that could compromise the validity and applicability of the simulated results in the context of electrophysiological data. Three scenarios involve microcircuits with 5 neurons, and one scenario involves a circuit with 20 neurons. Additionally, each scenario varies in terms of sparsity (number of connected neurons) and balance (number of excitatory and inhibitory neurons). In this sense, we define the following scenarios: 
\begin{itemize}
\item \textbf{Scenario 1:} The fixed synaptic weight matrix was constructed such that among the $20$ possible connections between neurons in the network {avoiding autapses}, we have the following distribution: $10\%$ of the connections are disconnected, meaning $2$ neurons have no connection between them. Additionally, $70\%$ of the connections are inhibitory ($14$ out of $20$ connections), while $20\%$ are excitatory ($4$ out of $20$ connections). The synaptic weight matrix, in this case, is given by
$$W = \begin{bmatrix}
0 & 0 & 1 & 1 & 1 \\
0 & 0 & 1 & 1 & 1 \\
1 & 1 & 0 & 1 & -4 \\
1 & 1 & 1 & 0 & -4 \\
1 & 1 & -4 & -4 & 0
\end{bmatrix}.$$

\item \textbf{Scenario 2:} The fixed synaptic weight matrix was constructed such that among the $20$ possible connections between neurons in the network {avoiding autapses}, we considered $10\%$ disconnected ($2$ neurons without a connection between them), $50\%$ inhibitory connections ($10$ out of $20$ connections), and $40\%$ excitatory connections ($8$ out of $20$ connections). The synaptic weight matrix, in this case, is given by
$$W = \begin{bmatrix}
0 & 0 & 3 & 3 & 3 \\
0 & 0 & 3 & 3 & 3 \\
3 & 3 & 0 & 3 & -12 \\
3 & 3 & 3 & 0 & -12 \\
3 & 3 & -12 & -12 & 0
\end{bmatrix}.$$

\item \textbf{Scenario 3:} The fixed synaptic weight matrix was constructed such that among the $20$ possible connections between neurons in the network {avoiding autapses}, we considered $10\%$ as disconnected ($2$ neurons without a connection between them), $30\%$ as inhibitory connections ($6$ out of $20$ connections), and $60\%$ as excitatory connections ($12$ out of $20$ connections). The synaptic weight matrix, in this case, is given by
$$W = \begin{bmatrix}
0 & 0 & 3 & 3 & 3 \\
0 & 0 & 1 & 1 & 1 \\
3 & 1 & 0 & 1 & -12 \\
3 & 1 & 1 & 0 & -4 \\
3 & 1 & -12 & -4 & 0
\end{bmatrix}.$$

\item \textbf{Scenario 4:} In this case, we fixed a circuit with $20$ neurons, resulting in a total of $380$ parameters to be estimated. We will consider $60\%$ of the $380$ possible connections as disconnections and maintain a ratio of $4$ excitatory neurons for every inhibitory neuron for the remaining connections, such that the excitatory connections have a synaptic weight of $4$ and the inhibitory connections have a synaptic weight of $-1$.
\end{itemize}

In all scenarios, we assess whether, as we increase the sample size, the estimates of the synaptic weights and the interaction neighborhoods become closer to the fixed values. In other words, we are interested in studying, computationally, the consistency of the estimators proposed in this paper. To this end, we fixed the following sample sizes: $T=500$, $T = 1000$, $T=5000$ and $T=10,000$. We also set four different values for the threshold $\epsilon$ that defines the interaction neighborhood estimator: $\epsilon = 10^{-5}$, $\epsilon = 10^{-4}$, $\epsilon = 10^{-3}$ and $\epsilon = 10^{-2}$.

\textcolor[rgb]{0,0,0}{The procedure for estimating interaction neighborhoods is based on a sensitivity measure that compares empirical firing probabilities. From a statistical inference perspective, given a postsynaptic neuron $i$, this measure can be interpreted as the test statistic used to assess, for each presynaptic neuron $j$, the null hypothesis that $j$ is not a neighbor of $i$, i.e., $\omega_{j \rightarrow i} = 0$. Using Pinsker's inequality \citep{pinsker1964information, kullback1967lower}, it can be shown that this statistic is upper-bounded by the likelihood ratio test statistic rescaled by the factor $1/T$. Consequently, at least asymptotically, the distribution of the sensitivity measure is dominated by a chi-square distribution. With this in mind, for the selection of the threshold $\epsilon$ in the simulation study, we first analyzed the behavior of the likelihood ratio test for different critical points of the form $T\epsilon$, obtained from various significance levels. The lower the significance level, the more extreme the critical points, reducing the probability of incorrectly rejecting the hypothesis that 
$j$ is not a neighbor of $i$ (Type I error). However, this also decreases the test's power, i.e., its ability to correctly identify true synaptic interactions. Thus, the values of $\epsilon$ were chosen to balance tolerance to Type I error and confidence in correct decisions, ensuring a statistically robust criterion for inferring interaction neighborhoods.}

\textcolor[rgb]{0,0,0}{By using the sensitivity measure $d$ instead of the likelihood ratio test statistic, the tendency is to estimate a less dense neuronal network, meaning fewer detected connections but with greater confidence that the identified connections are real. This occurs because $d$ is less sensitive than the likelihood ratio statistic, requiring stronger perturbations between neurons for a connection to be recognized. However, as aforementioned, experimental studies indicate that most neuronal connections in higher-order brain areas (e.g., neocortex) form sparse interaction graphs, which aligns with the biological plausibility of energy efficiency and noise minimization in neural circuits \citep{van2011rich, ercsey2013predictive}. This justifies the use of the measure $d$ to infer the interaction neighborhood, as this approach reduces the risk of including spurious connections that do not represent the true organization of the neural circuit, while still capturing the essential connectivity patterns.}

In each specified scenario, $100$ Monte Carlo replicas were generated for each sample size. For each replica, the synaptic weights and the connectivity graph were estimated using the methods considered in this study. The effectiveness of these methods in estimating the synaptic weights was evaluated based on the empirical mean squared error for networks with $5$ neurons and the average Euclidean distance between the estimated matrices and the original synaptic weight matrix for the network with $20$ neurons. The performance of the methods in estimating the connectivity graph was assessed by analyzing the proportions of correctly identified synaptic connections.

\subsection{Consistency of ML estimator for synaptic weight matrix}
\label{subsec62}

By analyzing Tables \ref{tab:MSE_MLE}, \ref{tab:MSE_MLE3} and \ref{tab:MSE_MLE5}, we observe that the empirical mean squared errors for each synaptic weight, calculated from $100$ Monte Carlo replicas, tend to decrease and approach zero as the sample sizes increase. There is not a single case where the error does not decrease as the sample size increases. 
Furthermore, Table \ref{tab:MSE_MLE7} shows that the Euclidean distance between the estimated synaptic weight matrix and the original matrix decreases on average as the sample size increases. Therefore, in all scenarios, the results indicate that the maximum likelihood (ML) estimators are consistent in estimating the synaptic weights for the considered neuronal model, as promised by Theorem \ref{thm:MLE_consistency}.

\begin{figure}[http!]
\centering
\includegraphics[scale=.2]{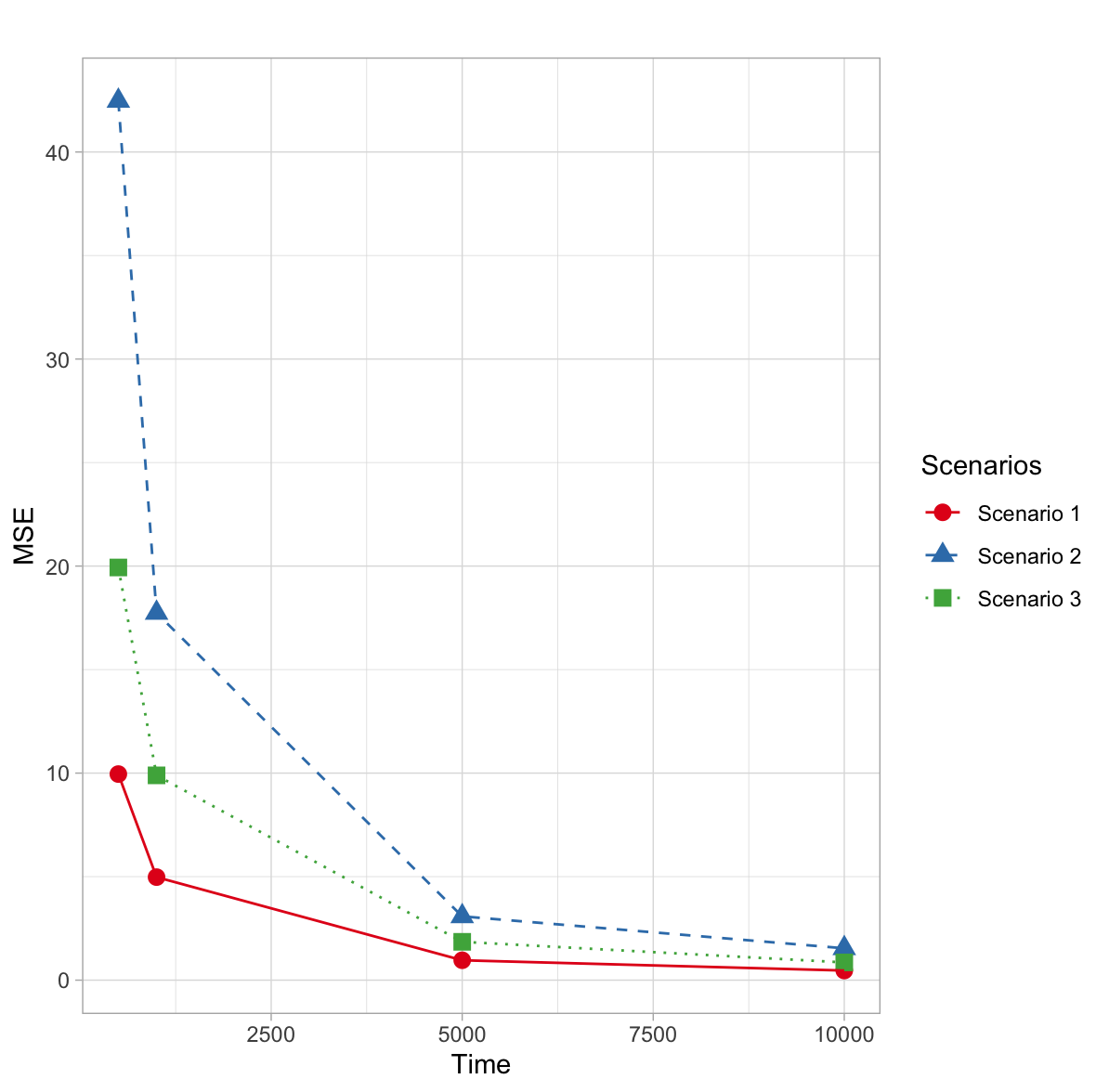}
\caption{Average Euclidean distance between estimated synaptic weight matrix and the original matrix from estimates of neuronal connectivity matrix of a network with 20 neurons, using the maximum likelihood method. The calculations were performed considering the first three scenarios and $4$ different sample sizes: $T = 500$, $T = 1000$, $T = 5000$, and $T = 10,000$.}
\label{ref:fig1}
\end{figure}

In Figure~\ref{ref:fig1}, we present the behavior of the average Euclidean distance between the estimated synaptic weight matrix and the original matrix across the first three scenarios. We observe that scenario 1, in which the connections are weaker, exhibits the smallest average distances for all sample sizes, followed by scenario 3, where there is a mix of strong and weak connections, and finally, scenario 2, where the connections are stronger. These results are expected. In scenarios with weaker connections, small variations in the data are likely to have less impact on the estimates of the synaptic weights, resulting in smaller Euclidean distances between the estimated and original matrices. In contrast, in scenarios with stronger connections, variations in the data may cause larger deviations in the estimates, leading to greater average Euclidean distances. The intermediate scenario, with a mix of strong and weak connections, naturally shows behavior that falls between these two extremes. 

Furthermore, we can observe that, compared to the other scenarios, in scenario 4, where we have a network with a larger number of neurons, the average Euclidean distance between the estimated synaptic weight matrix and the original matrix is considerably higher for all sample sizes. These results are expected. With more connections to estimate, the variability in synaptic weights tends to be higher, especially for smaller sample sizes, resulting in greater Euclidean distances between the estimated and original matrices. Even as the sample size increases, capturing the full structure of a larger network remains more difficult than in smaller networks, which explains why the average distance is consistently higher in this scenario.

\subsection{Consistency of interaction neighborhood estimation procedure}
\label{subsec63}

From the analysis of the results in Tables \ref{fig01:Scenario01_Model_Selection}, \ref{fig02:Scenario02_Model_Selection}, \ref{fig03:Scenario03_Model_Selection} and \ref{tab9:model_selection_scenario4}, we observe that, as desired and expected, smaller cutoff values $\epsilon$ lead to a higher proportion of false positives (disconnections incorrectly identified), while larger $\epsilon$ values result in a higher proportion of false negatives (connections incorrectly identified). Furthermore, we note that the proportion of times the presence or absence of synaptic connections is correctly identified tends to increase as the sample size grows for suitable cutoff values. This indicates that the proposed method is consistent in estimating interaction neighborhoods, as promised by Theorem \ref{thm:model_selection_consistency}.

An analysis of the results obtained with the microcircuits reveals that, regardless of the strength of excitatory and inhibitory synapses within the network, the methods face greater challenges in estimating synaptic weights when the sample size is small. However, these challenges are progressively mitigated as the sample size increases. This pattern is also observed in the sparse network with $20$ neurons. In practice, cortical neurons typically fire within $1$ to $3$ milliseconds (ms), and a bin size of $0.30$ ms is commonly used for temporal resolution (see \citealp{softky1993highly}). The low empirical mean squared errors observed for $T = 10,000$ suggest that in real-world scenarios, where sample sizes are often much larger, the method performs well in accurately estimating both synaptic weights and connectivity graphs. Thus, we conclude that the proposed methodology is robust in terms of both network imbalance and sparsity. 

\begin{table}[http!]
    {\footnotesize\caption{\textbf{Scenario 1.} Empirical mean squared error calculated for Scenario $1$ from $100$ estimates of the neuronal connectivity matrix of a network with $5$ neurons, using the maximum likelihood method. The calculations were performed considering $4$ different sample sizes: $T=500$, $T=1000$, $T=5000$, $T=10,000$.}
    \begin{tabular}{c|c|c|c|c|c}
    \toprule
        \textbf{Synaptic weights} & \textbf{Values} & $\boldsymbol{T = 500}$ & $\boldsymbol{T = 1000}$ & $\boldsymbol{T = 5000}$ & $\boldsymbol{T = 10,000}$  \\
    \midrule
        $\omega_{2 \rightarrow 1}$ & $0$ & $0.4136$ & $0.2276$ & $0.0309$ & $0.0187$\\
        $\omega_{3 \rightarrow 1}$ & $1$ & $0.3570$ & $0.2250$ & $0.0415$ & $0.0175$\\
        $\omega_{4 \rightarrow 1}$ & $1$ & $0.3352$ & $0.1711$ & $0.0351$ & $0.0166$\\
        $\omega_{5 \rightarrow 1}$ & $1$ & $0.4414$ & $0.2549$ & $0.0399$ & $0.0246$\\
        $\omega_{1 \rightarrow 2}$ & $0$ & $0.4014$ & $0.1807$ & $0.0369$ & $0.0187$\\
        $\omega_{3 \rightarrow 2}$ & $1$ & $0.3808$ & $0.1574$ & $0.0425$ & $0.0132$\\
        $\omega_{4 \rightarrow 2}$ & $1$ & $0.3385$ & $0.1834$ & $0.0359$ & $0.0189$\\
        $\omega_{5 \rightarrow 2}$ & $1$ & $0.5847$ & $0.2226$ & $0.0486$ & $0.0186$\\
        $\omega_{1 \rightarrow 3}$ & $1$ & $0.4479$ & $0.2571$ & $0.0472$ & $0.0210$\\
        $\omega_{2 \rightarrow 3}$ & $1$ & $0.4068$ & $0.2204$ & $0.0401$ & $0.0196$\\
        $\omega_{4 \rightarrow 3}$ & $1$ & $0.4631$ & $0.2265$ & $0.0386$ & $0.0248$\\
        $\omega_{5 \rightarrow 3}$ & $-4$ & $0.3896$ & $0.2941$ & $0.0500$ & $0.0242$\\
        $\omega_{1 \rightarrow 4}$ & $1$ & $0.3857$ & $0.1615$ & $0.0348$ & $0.0203$\\
        $\omega_{2 \rightarrow 4}$ & $1$ & $0.4514$ & $0.2294$ & $0.0387$ & $0.0196$\\
        $\omega_{3 \rightarrow 4}$ & $1$ & $0.4211$ & $0.2267$ & $0.0366$ & $0.0239$\\
        $\omega_{5 \rightarrow 4}$ & $-4$ & $0.5247$ & $0.2695$ & $0.0499$ & $0.0266$\\
        $\omega_{1 \rightarrow 5}$ & $1$ & $0.6469$ & $0.2816$ & $0.0716$ & $0.0287$\\
        $\omega_{2 \rightarrow 5}$ & $1$ & $0.6381$ & $0.3196$ & $0.0727$ & $0.0273$\\
        $\omega_{3 \rightarrow 5}$ & $-4$ & $1.1018$ & $0.4617$ & $0.0848$ & $0.0418$\\
        $\omega_{4 \rightarrow 5}$ & $-4$ & $0.8240$ & $0.4049$ & $0.0844$ & $0.0392$\\
    \bottomrule
    \end{tabular}
    \label{tab:MSE_MLE}}
\end{table}

\begin{table}[http!]
    {\footnotesize\caption{\textbf{Scenario 2.} Empirical mean squared error calculated for Scenario $2$ from $100$ estimates of the neuronal connectivity matrix of a network with $5$ neurons, using the maximum likelihood method. The calculations were performed considering $4$ different sample sizes: $T=500$, $T=1000$, $T=5000$, $T=10,000$.}
    \begin{tabular}{c|c|c|c|c|c}
    \toprule
        \textbf{Synaptic weights} & \textbf{Values} & $\boldsymbol{T = 500}$ & $\boldsymbol{T = 1000}$ & $\boldsymbol{T = 5000}$ & $\boldsymbol{T = 10,000}$  \\
    \midrule
        $\omega_{2 \rightarrow 1}$ & $0$ & $0.6172$ & $0.2395$ & $0.0650$ & $0.0370$\\
        $\omega_{3 \rightarrow 1}$ & $3$ & $0.8812$ & $0.5853$ & $0.0752$ & $0.0459$\\
        $\omega_{4 \rightarrow 1}$ & $3$ & $0.8360$ & $0.3529$ & $0.0901$ & $0.0407$\\
        $\omega_{5 \rightarrow 1}$ & $3$ & $2.1359$ & $0.5699$ & $0.1201$ & $0.0576$\\
        $\omega_{1 \rightarrow 2}$ & $0$ & $0.4270$ & $0.2447$ & $0.0528$ & $0.0296$\\
        $\omega_{3 \rightarrow 2}$ & $3$ & $1.0395$ & $0.5607$ & $0.0871$ & $0.0349$\\
        $\omega_{4 \rightarrow 2}$ & $3$ & $0.7968$ & $0.4588$ & $0.0847$ & $0.0366$\\
        $\omega_{5 \rightarrow 2}$ & $3$ & $1.3990$ & $0.4803$ & $0.1161$ & $0.0613$\\
        $\omega_{1 \rightarrow 3}$ & $3$ & $1.4537$ & $0.7800$ & $0.1052$ & $0.0542$\\
        $\omega_{2 \rightarrow 3}$ & $3$ & $1.2568$ & $0.7117$ & $0.1338$ & $0.0575$\\
        $\omega_{4 \rightarrow 3}$ & $3$ & $1.0548$ & $0.4849$ & $0.1189$ & $0.0484$\\
        $\omega_{5 \rightarrow 3}$ & $-12$ & $2.6229$ & $1.6650$ & $0.2498$ & $0.1413$\\
        $\omega_{1 \rightarrow 4}$ & $3$ & $1.6610$ & $0.7104$ & $0.0879$ & $0.0525$\\
        $\omega_{2 \rightarrow 4}$ & $3$ & $2.0422$ & $0.8964$ & $0.1026$ & $0.0630$\\
        $\omega_{3 \rightarrow 4}$ & $3$ & $1.0559$ & $0.5947$ & $0.1240$ & $0.0575$\\
        $\omega_{5 \rightarrow 4}$ & $-12$ & $4.0172$ & $1.5245$ & $0.2356$ & $0.1129$\\
        $\omega_{1 \rightarrow 5}$ & $3$ & $2.9085$ & $0.8160$ & $0.1748$ & $0.0737$\\
        $\omega_{2 \rightarrow 5}$ & $3$ & $2.2852$ & $0.9990$ & $0.1559$ & $0.0667$\\
        $\omega_{3 \rightarrow 5}$ & $-12$ & $7.2263$ & $2.0596$ & $0.5007$ & $0.2413$\\
        $\omega_{4 \rightarrow 5}$ & $-12$ & $6.7505$ & $3.0026$ & $0.4006$ & $0.2255$\\
    \bottomrule
    \end{tabular}
    \label{tab:MSE_MLE3}}
\end{table}

\begin{table}[http!]
    {\footnotesize\caption{\textbf{Scenario 3.} Empirical mean squared error calculated for Scenario $3$ from $100$ estimates of the neuronal connectivity matrix of a network with $5$ neurons, using the maximum likelihood method. The calculations were performed considering $4$ different sample sizes: $T=500$, $T=1000$, $T=5000$, $T=10,000$.}
    \begin{tabular}{c|c|c|c|c|c}
    \toprule
        \textbf{Synaptic weights} & \textbf{Values} & $\boldsymbol{T = 500}$ & $\boldsymbol{T = 1000}$ & $\boldsymbol{T = 5000}$ & $\boldsymbol{T = 10,000}$  \\
    \midrule
        $\omega_{2 \rightarrow 1}$ & $0$ & $0.7128$ & $0.3135$ & $0.0668$ & $0.0325$\\
        $\omega_{3 \rightarrow 1}$ & $3$ & $1.1546$ & $0.7102$ & $0.0725$ & $0.0356$\\
        $\omega_{4 \rightarrow 1}$ & $3$ & $0.8493$ & $0.4073$ & $0.0922$ & $0.0334$\\
        $\omega_{5 \rightarrow 1}$ & $3$ & $1.2718$ & $0.5657$ & $0.1488$ & $0.0626$\\
        $\omega_{1 \rightarrow 2}$ & $0$ & $0.3418$ & $0.1485$ & $0.0318$ & $0.0173$\\
        $\omega_{3 \rightarrow 2}$ & $1$ & $0.3340$ & $0.2000$ & $0.0520$ & $0.0158$\\
        $\omega_{4 \rightarrow 2}$ & $1$ & $0.3253$ & $0.1657$ & $0.0429$ & $0.0152$\\
        $\omega_{5 \rightarrow 2}$ & $1$ & $0.6326$ & $0.2358$ & $0.0450$ & $0.0267$\\
        $\omega_{1 \rightarrow 3}$ & $3$ & $0.8840$ & $0.5797$ & $0.0823$ & $0.0354$\\
        $\omega_{2 \rightarrow 3}$ & $1$ & $0.8849$ & $0.3842$ & $0.0796$ & $0.0414$\\
        $\omega_{4 \rightarrow 3}$ & $1$ & $0.6883$ & $0.3270$ & $0.0885$ & $0.0352$\\
        $\omega_{5 \rightarrow 3}$ & $-12$ & $2.9491$ & $1.7657$ & $0.2977$ & $0.1148$\\
        $\omega_{1 \rightarrow 4}$ & $3$ & $0.6040$ & $0.2842$ & $0.0444$ & $0.0271$\\
        $\omega_{2 \rightarrow 4}$ & $1$ & $0.5841$ & $0.2662$ & $0.0454$ & $0.0239$\\
        $\omega_{3 \rightarrow 4}$ & $1$ & $0.4160$ & $0.2163$ & $0.0503$ & $0.0217$\\
        $\omega_{5 \rightarrow 4}$ & $-4$ & $0.6854$ & $0.3433$ & $0.0562$ & $0.0284$\\
        $\omega_{1 \rightarrow 5}$ & $3$ & $0.9871$ & $0.5951$ & $0.0913$ & $0.0429$\\
        $\omega_{2 \rightarrow 5}$ & $1$ & $0.7553$ & $0.4046$ & $0.0801$ & $0.0364$\\
        $\omega_{3 \rightarrow 5}$ & $-12$ & $3.5802$ & $1.5775$ & $0.2970$ & $0.1551$\\
        $\omega_{4 \rightarrow 5}$ & $-4$ & $1.2914$ & $0.4006$ & $0.0859$ & $0.0546$\\
    \bottomrule
    \end{tabular}
    \label{tab:MSE_MLE5}}
\end{table}

\begin{table}[http!]
    {\footnotesize\caption{\textbf{Scenario 4.} Average Euclidean distance between the estimated synaptic weight matrix and the original matrix for Scenario $4$ from $100$ estimates of the neuronal connectivity matrix of a network with $20$ neurons, using the maximum likelihood method. The calculations were performed considering $4$ different sample sizes: $T=500$, $T=1000$, $T=5000$, $T=10,000$.}
    \begin{tabular}{c|c|c|c|c|c}
    \toprule
        \textbf{Synaptic matrix} & \textbf{Values} & $\boldsymbol{T = 500}$ & $\boldsymbol{T = 1000}$ & $\boldsymbol{T = 5000}$ & $\boldsymbol{T = 10,000}$  \\
    \midrule
    $W$ & See Subsection & $2349.749$ & $625.0891$ & $94.9156$ & $45.6833$\\
    \bottomrule
    \end{tabular}
    \label{tab:MSE_MLE7}}
\end{table}

\begin{table}[http!]
    \centering
    \caption{\textbf{Scenario 1.} Identification of the presence and absence of connections in a network with 5 neurons for Scenario 1. The proportions of correctly identified synaptic connections is calculated considering 100 Monte Carlo replicas. The calculations were performed considering four different sample sizes ($T=500$, $T=1,000$, $T=5,000$, $T=10,000$) and four different cuttof values ($\epsilon = 10^{-5}$, $\epsilon = 10^{-4}$, $\epsilon = 10^{-3}$, $\epsilon = 10^{-2}$).}
    \includegraphics[width=1.01\linewidth]{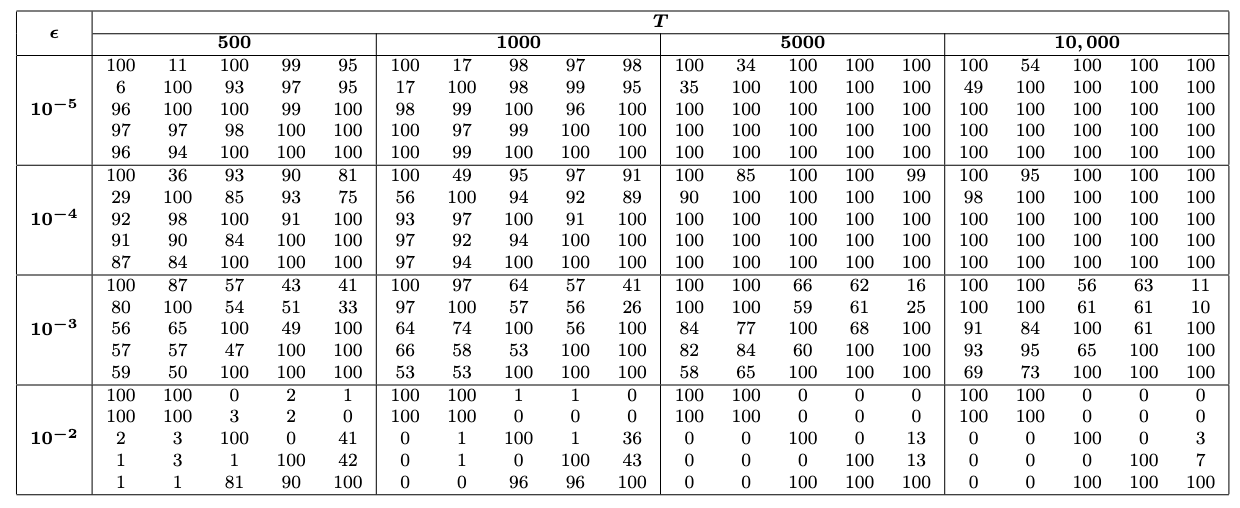}
    \label{fig01:Scenario01_Model_Selection}
\end{table}

\begin{table}[http!]
    \centering
    \caption{\textbf{Scenario 2.} Identification of the presence and absence of connections in a network with 5 neurons for Scenario 2. The proportions of correctly identified synaptic connections is calculated considering 100 Monte Carlo replicas. The calculations were performed considering four different sample sizes ($T=500$, $T=1,000$, $T=5,000$, $T=10,000$) and four different cuttof values ($\epsilon = 10^{-5}$, $\epsilon = 10^{-4}$, $\epsilon = 10^{-3}$, $\epsilon = 10^{-2}$).}
    \includegraphics[width=1.01\linewidth]{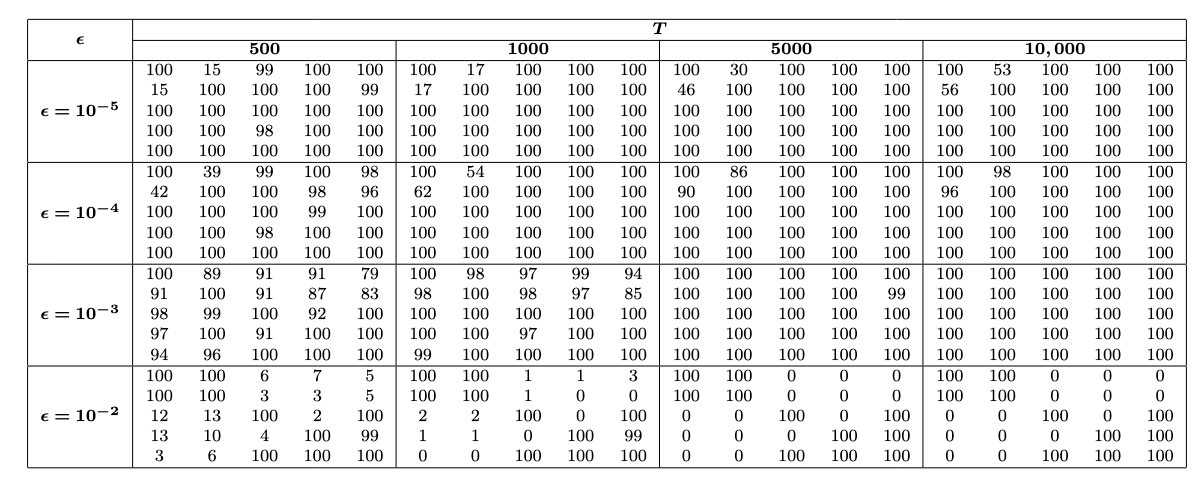}
    \label{fig02:Scenario02_Model_Selection}
\end{table}

\newpage

\begin{table}[http!]
    \centering
    \caption{\textbf{Scenario 3.} Identification of the presence and absence of connections in a network with 5 neurons for Scenario 3. The proportions of correctly identified synaptic connections is calculated considering 100 Monte Carlo replicas. The calculations were performed considering four different sample sizes ($T=500$, $T=1,000$, $T=5,000$, $T=10,000$) and four different cuttof values ($\epsilon = 10^{-5}$, $\epsilon = 10^{-4}$, $\epsilon = 10^{-3}$, $\epsilon = 10^{-2}$).}
    \includegraphics[width=1.01\linewidth]{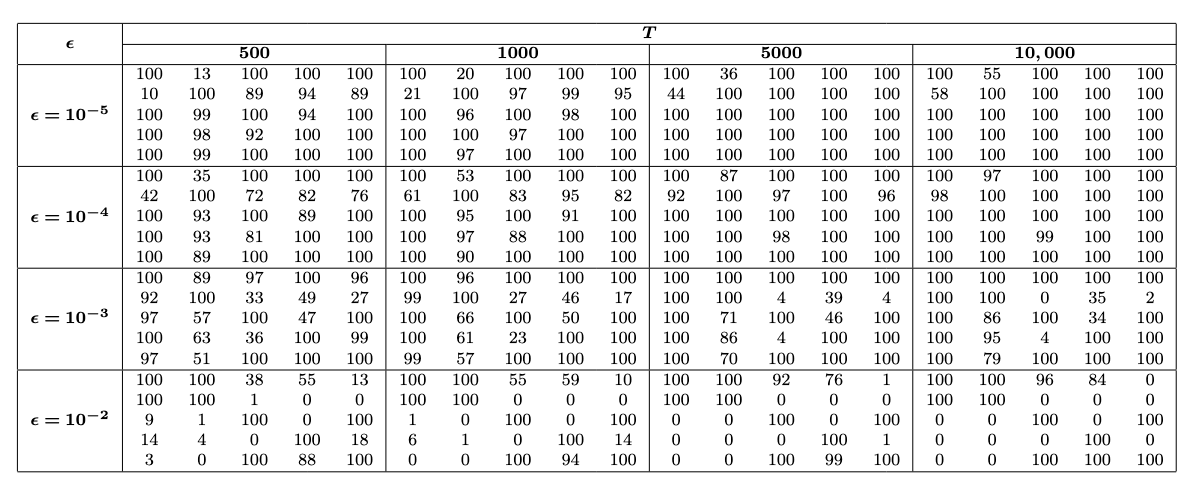}
    \label{fig03:Scenario03_Model_Selection}
\end{table}

\begin{table}[http!]
\centering
 \caption{\textbf{Scenario 4.} Identification of the presence and absence of connections in a network with 20 neurons for Scenario 4. The proportions of correctly identified synaptic connections is calculated considering 100 Monte Carlo replicas. The calculations were performed considering four different sample sizes ($T=500$, $T=1,000$, $T=5,000$, $T=10,000$) and four different cuttof values ($\epsilon = 10^{-5}$, $\epsilon = 10^{-4}$, $\epsilon = 10^{-3}$, $\epsilon = 10^{-2}$).}
\begin{tabular}{|c|c|c|c|c|}
\hline
                        & \multicolumn{4}{|c|}{$\boldsymbol{T}$}                                                   \\
                        \hline
$\boldsymbol{\epsilon}$ & $\boldsymbol{500}$ & $\boldsymbol{1000}$ & $\boldsymbol{5000}$ & $\boldsymbol{10,000}$ \\
\hline
$\boldsymbol{10^{-5}}$  & $0.5712$           & $0.5869$            & $0.6512$            & $0.7062$              \\
\hline
$\boldsymbol{10^{-4}}$  & $0.6118$           & $0.6549$            & $0.7566$            & $0.7754$              \\
\hline
$\boldsymbol{10^{-3}}$  & $0.6586$           & $0.6866$            & $0.7173$            & $0.7266$              \\
\hline
$\boldsymbol{10^{-2}}$  & $0.5682$           & $0.5711$            & $0.5756$            & $0.5763$   \\
\hline
\end{tabular}
\label{tab9:model_selection_scenario4}
\end{table}

\section{Application to neurobiological data}
\label{sec8}

\textcolor[rgb]{0,0,0}{In this section, we illustrate the usefulness of the proposed method in an experimental data set. Understanding the connectivity within neural circuits is essential for drawing principles governing brain function and dynamics. Through computational modeling, we translate noisy multi-unit data into a structured connectivity matrix, enabling the simulation of dynamic neural circuits and offering a platform for deeper investigation into the interactions and behaviors that emerge within these networks. }

\textcolor[rgb]{0,0,0}{In applying our model selection procedure to electrophysiological data from the rat hippocampus, we first transformed raw neuronal firing timestamps into binary spike trains using a bin size of 1 ms to preserve temporal resolution. The choice of bin size was informed by the characteristic firing frequency of CA1 neurons. Prior to analysis, we performed artifact removal and quality control measures to ensure data integrity. The resulting connectivity matrices revealed clear inhibitory and excitatory relationships consistent with known CA1 circuitry, thus validating the biological relevance of our inferred network interactions.}

\textcolor[rgb]{0,0,0}{More specifically, we select 5 traces from multichannel simultaneous recordings made from the CA1 of rats, data freely available at \url{https://crcns.org/data-sets/hc} where several other neurons can be found. The neurons can be either pyramidal cells or interneurons, i.e., excitatory or inhibitory, respectively. The database we use is composed of vectors that record the firing moments of each neuron. Before training the model, it is necessary to transform these time markings into spike-train vectors indicating whether or not there is a firing at a specific time $t$ for a given neuron $i$. We then convert the data into a binary matrix (0 or 1), where the rows represent the neurons and the columns represent the time intervals, indicating the occurrence of firings. Thus, at the end of the process, we obtain a sample of $T = 2$s for all 4 neurons. A more careful analysis of this ensemble of neurons reveals that they fire at 7.14 Hz, 9.97 Hz, 6.71 Hz, 9.97 Hz, and 8.12 Hz, which are typical firing rates for neurons in the CA1 area.}

\begin{figure}[http!]
    \centering
    \includegraphics[width=0.7\linewidth]{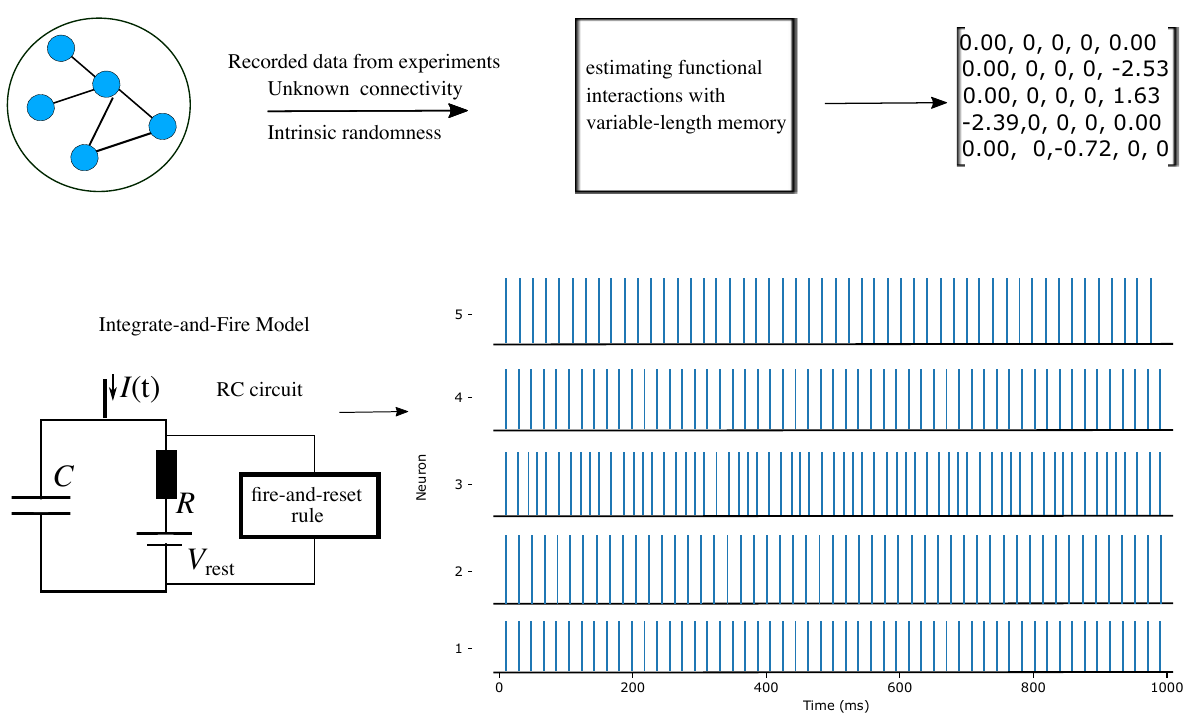}
    \caption{{\textbf{Connectivity Extraction Workflow}. Electrophysiological recordings from a specific brain area, with unknown connectivity, serve as the initial data source. Due to the stochastic nature of the neural environment, the recorded data is inherently noisy. A connectivity matrix estimating functional interactions is then derived from the data, constrained to a range between -10 and 10 in this example. This matrix can subsequently inform simulations of microcircuits composed of computational models. In the figure's bottom section, we demonstrate a simulation using five leaky integrate-and-fire neuron models with a threshold of -50 mV and a reset potential of -65 mV, along with a membrane time constant of 10 ms. Connections are conductance-based and categorized as either excitatory or inhibitory, with respective time constants of 0.5 ms and reversal potentials of 0 mV or -70 mV. Conductance values are taken directly from the estimated matrix. The resulting neuronal firing activity offers a basis for further analysis and experimentation.}}
    \label{fig:application}
\end{figure}

{Figure~\ref{fig:application} illustrates the workflow applied for extracting and utilizing connectivity data from the electrophysiological recordings to simulate neural microcircuits. Initially, noisy electrophysiological recordings from a specific brain region with unknown connectivity are processed. Through this workflow, a connectivity matrix is derived, estimating the functional interactions between neurons. We use results from $T = 19,999$ with $\epsilon = 10^{-4}$ to build this connectivity matrix. In this example, the matrix captures both inhibitory and excitatory connections. The connectivity matrix is then applied to simulate a microcircuit comprising five leaky integrate-and-fire neuron models. Each neuron has a firing threshold of -50 mV, a reset potential of -65 mV, and a membrane time constant of 10 ms. Synaptic connections are modeled as conductance-based inputs, designated as excitatory or inhibitory with respective time constants of 0.5 ms and reversal potentials of 0 mV or -70 mV. These conductance values are drawn directly from the estimated connectivity matrix, enabling the simulation of realistic neuronal firing activity. }

{This framework supports further experimentation and analysis, providing insights into the functional connectivity and dynamics of neuronal networks. Future analysis could include more specific neurons with ion currents, i.e. distinguishing between pyramidal cells or interneurons. The code for the simulation in Fig.~\ref{fig:application} is freely available at \url{https://github.com/rodrigo-pena-lab/functional_interactions}.}

\section{Proofs}
\label{sec9}

In this section, we provide the proofs of Theorems \ref{thm:MLE_consistency} and \ref{thm:model_selection_consistency}.

\subsection{Proof of Theorem \ref{thm:MLE_consistency}} 

We begin with the proof of Theorem \ref{thm:MLE_consistency}. First, we present a proof for the strong consistency of the maximum likelihood estimators for the synaptic weights, followed by a proof for the $L^2$ consistency.
  
\subsubsection{Strong Consistency}

For each $\boldsymbol{u} \in \mathcal{T}^{(i)}_T$, we define the row vector 
$$\hat{Q}_{\boldsymbol{u}}^{(i)} := \left(\dfrac{N_{T}^{(i)}(\boldsymbol{u}, 0)}{N_{T}^{(i)}(\boldsymbol{u})}; \ \dfrac{N_{T}^{(i)}(\boldsymbol{u}, 1)}{N_{T}^{(i)}(\boldsymbol{u})}\right) \in M_{1 \times 2}\left([0,1]\right),$$ 
which is the empirical distribution conditioned on the configuration $\boldsymbol{u}$.  In equality (1), we define $P_{\boldsymbol{u}a}(\boldsymbol{\omega}^{(i)})$ as the transition probability from $\boldsymbol{u} \in \mathcal{T}_{T}^{(i)}$ to $a \in \{0,1\}$ according to (3.3). In this context, define the transition distribution out of configuration $\boldsymbol{u}$ as a row vector 
$$P_{\boldsymbol{u}}(\boldsymbol{\omega}^{(i)}) := \left(P_{\boldsymbol{u}0}(\boldsymbol{\omega}^{(i)}); \ P_{\boldsymbol{u}1}(\boldsymbol{\omega}^{(i)})\right) \in M_{1 \times 2}\left([0,1]\right),$$
which is a row in the transition matrix $P(\boldsymbol{\omega}^{(i)}) := (P_{\boldsymbol{u}a}(\boldsymbol{\omega}^{(i)}))$

It is well-known that the set $\left\{\hat{Q}^{(i)}_{\boldsymbol{u}}: \boldsymbol{u} \in \mathcal{T}_{T}^{(i)}\right\}$ is the ML estimator of the transition matrix $P := (P_{\boldsymbol{u}a})$, assuming no further parametrization of the transition probabilities (see, for example, \citealp{buhlmann1999variable}). Consider the ML estimator $\hat{\boldsymbol{\omega}}_T^{(i)}$ for $\boldsymbol{\omega}^{(i)}$ defined in (3) and define the row vector  
$$P_{\boldsymbol{u}}(\hat{\boldsymbol{\omega}}^{(i)}_{T}) := \left(P_{\boldsymbol{u}0}(\hat{\boldsymbol{\omega}}^{(i)}_{T}); \ P_{\boldsymbol{u}1}(\hat{\boldsymbol{\omega}}^{(i)}_{T})\right) \in M_{1 \times 2}\left([0,1]\right),$$
as an estimator of $P_{\boldsymbol{u}}(\boldsymbol{\omega}^{(i)})$. In this sense, consider the Kullback-Leibler distance between the estimators $\hat{Q}_{\boldsymbol{u}}^{(i)}$ and $P_{\boldsymbol{u}}(\boldsymbol{\omega}^{(i)})$:
$$D_{KL}\left(\hat{Q}_{\boldsymbol{u}}^{(i)} \left\| P_{\boldsymbol{u}}(\hat{\boldsymbol{\omega}}^{(i)}_T) \right.\right) := - \sum_{a = 0}^{1} \dfrac{N_{T}^{(i)}(\boldsymbol{u}, a)}{N_{T}^{(i)}(\boldsymbol{u})} \log\left(\dfrac{P_{\boldsymbol{u}a}(\hat{\boldsymbol{\omega}}^{(i)}_T)}{{N_{T}^{(i)}(\boldsymbol{u}, a)}/{N_{T}^{(i)}(\boldsymbol{u})}}\right).$$
By the non-negativity of Kullback-Leibler distance, we have
$$\sum_{a = 0}^{1} \dfrac{N_{T}^{(i)}(\boldsymbol{u}, a)}{N_{T}^{(i)}(\boldsymbol{u})} \log\left(\dfrac{N_{T}^{(i)}(\boldsymbol{u}, a)}{N_{T}^{(i)}(\boldsymbol{u})}\right) \geq \sum_{a = 0}^{1}  \dfrac{N_{T}^{(i)}(\boldsymbol{u}, a)}{N_{T}^{(i)}(\boldsymbol{u})} \log\left(P_{\boldsymbol{u}a}(\hat{\boldsymbol{\omega}}^{(i)}_T)\right),$$
which implies that
\begin{equation}
\sum_{\boldsymbol{u} \in \mathcal{T}^{(i)}_T}\sum_{a = 0}^{1} \dfrac{N_{T}^{(i)}(\boldsymbol{u}, a)}{T} \log\left(\dfrac{N_{T}^{(i)}(\boldsymbol{u}, a)}{N_{T}^{(i)}(\boldsymbol{u})}\right) \geq \sum_{\boldsymbol{u} \in \mathcal{T}^{(i)}_T} \sum_{a = 0}^{1}  \dfrac{N_{T}^{(i)}(\boldsymbol{u}, a)}{T} \log\left(P_{\boldsymbol{u}a}(\hat{\boldsymbol{\omega}}^{(i)}_T)\right).
\label{eq1:proof_thm1}
\end{equation}
Since $\hat{\boldsymbol{\omega}}^{(i)}_T$ is the ML estimator of $\boldsymbol{\omega}^{(i)}$ follows, by assumption \eqref{assumption3:spike_rate_function_bounded}, that 
\begin{equation}
\sum_{\boldsymbol{u} \in \mathcal{T}^{(i)}_T} \sum_{a = 0}^{1}  \dfrac{N_{T}^{(i)}(\boldsymbol{u}, a)}{T} \log\left(P_{\boldsymbol{u}a}(\hat{\boldsymbol{\omega}}^{(i)}_T)\right) \geq \sum_{\boldsymbol{u} \in \mathcal{T}^{(i)}_T} \sum_{a = 0}^{1}  \dfrac{N_{T}^{(i)}(\boldsymbol{u}, a)}{T} \log\left(P_{\boldsymbol{u}a}({\boldsymbol{\omega}}^{(i)}_0)\right). 
\label{eq2:proof_thm1}
\end{equation}

By the Ergodic Theorem, almost surely, 
\begin{equation}
\lim_{T \rightarrow \infty} \sum_{\boldsymbol{u} \in \mathcal{T}^{(i)}_T}\sum_{a = 0}^{1} \dfrac{N_{T}^{(i)}(\boldsymbol{u}, a)}{T} \log\left(\dfrac{N_{T}^{(i)}(\boldsymbol{u}, a)}{N_{T}^{(i)}(\boldsymbol{u})}\right) = \sum_{\boldsymbol{u}} \sum_{a = 0}^{1} \pi_{\boldsymbol{u}}(\boldsymbol{\omega}^{(i)}_0) P_{\boldsymbol{u}a}(\boldsymbol{\omega}^{(i)}_0) \log P_{\boldsymbol{u}a}(\boldsymbol{\omega}^{(i)}_0),
\label{eq:ergodic_theorem_81}
\end{equation}
and also, almost surely,
\begin{equation}
\lim_{T \rightarrow \infty}\sum_{\boldsymbol{u} \in \mathcal{T}^{(i)}_T} \sum_{a = 0}^{1}  \dfrac{N_{T}^{(i)}(\boldsymbol{u}, a)}{T} \log\left(P_{\boldsymbol{u}a}({\boldsymbol{\omega}}^{(i)}_0)\right) = \sum_{\boldsymbol{u}} \sum_{a = 0}^{1} \pi_{\boldsymbol{u}}(\boldsymbol{\omega}^{(i)}_0) P_{\boldsymbol{u}a}(\boldsymbol{\omega}^{(i)}_0) \log P_{\boldsymbol{u}a}(\boldsymbol{\omega}^{(i)}_0),
\label{eq:ergodic_theorem_82}
\end{equation}

Subtracting
$$\sum_{\boldsymbol{u} \in \mathcal{T}^{(i)}_T}\sum_{a = 0}^{1} \dfrac{N_{T}^{(i)}(\boldsymbol{u}, a)}{T} \log\left(\dfrac{N_{T}^{(i)}(\boldsymbol{u}, a)}{N_{T}^{(i)}(\boldsymbol{u})}\right)$$
from inequalities \eqref{eq1:proof_thm1} and \eqref{eq2:proof_thm1} and then combining them, we obtain
\begin{align}
&\sum_{\boldsymbol{u} \in \mathcal{T}^{(i)}_T} \sum_{a = 0}^{1}  \dfrac{N_{T}^{(i)}(\boldsymbol{u}, a)}{T} \log\left(\dfrac{P_{\boldsymbol{u}a}({\boldsymbol{\omega}}^{(i)}_0)}{N_{T}^{(i)}(\boldsymbol{u}, a)/ N_T^{(i)}(\boldsymbol{u})}\right) \nonumber \\
&\leq \sum_{\boldsymbol{u} \in \mathcal{T}^{(i)}_T} \sum_{a = 0}^{1}  \dfrac{N_{T}^{(i)}(\boldsymbol{u}, a)}{T} \log\left(\dfrac{P_{\boldsymbol{u}a}(\hat{\boldsymbol{\omega}}^{(i)}_T)}{N_{T}^{(i)}(\boldsymbol{u}, a)/ N_T^{(i)}(\boldsymbol{u})}\right) \nonumber \\
&\leq 0.
\label{eq:inequality_85}
\end{align}

By \eqref{eq:ergodic_theorem_81} and \eqref{eq:ergodic_theorem_82}, we have that, almost surely,
$$\lim_{T \rightarrow \infty} \sum_{\boldsymbol{u} \in \mathcal{T}^{(i)}_T} \sum_{a = 0}^{1}  \dfrac{N_{T}^{(i)}(\boldsymbol{u}, a)}{T} \log\left(\dfrac{P_{\boldsymbol{u}a}({\boldsymbol{\omega}}^{(i)}_0)}{N_{T}^{(i)}(\boldsymbol{u}, a)/ N_T^{(i)}(\boldsymbol{u})}\right) = 0,$$
Therefore, by \eqref{eq:inequality_85}, we conclude that, almost surely,
\begin{equation}
\lim_{T \rightarrow \infty} \sum_{\boldsymbol{u} \in \mathcal{T}^{(i)}_T} \sum_{a = 0}^{1}  \dfrac{N_{T}^{(i)}(\boldsymbol{u}, a)}{T} \log\left(\dfrac{P_{\boldsymbol{u}a}(\hat{\boldsymbol{\omega}}^{(i)}_T)}{N_{T}^{(i)}(\boldsymbol{u}, a)/ N_T^{(i)}(\boldsymbol{u})}\right) = 0.
\label{eq:ergodic_theorem_86}
\end{equation}

By Pinsker's inequality (see, for example, \citealp{pinsker1964information, kullback1967lower}), 
$$\left\|\hat{Q}_{\boldsymbol{u}}^{(i)} - P_{\boldsymbol{u}}(\hat{\boldsymbol{\omega}}^{(i)})\right\|_2^2 \leq 2 D_{KL}\left(\hat{Q}^{(i)}_{\boldsymbol{u}} \left\| P_{\boldsymbol{u}}(\hat{\boldsymbol{\omega}}^{(i)}) \right.\right),$$
which implies that
\begin{align}
0 &\leq \sum_{\boldsymbol{u} \in \mathcal{T}_T^{(i)}} \sum_{a = 0}^{1} \dfrac{N_{T}^{(i)}(\boldsymbol{u})}{T} \left(P_{\boldsymbol{u}a}(\hat{\boldsymbol{\omega}}^{(i)}_T) - \dfrac{N_{T}^{(i)}(\boldsymbol{u}, a)}{N_{T}^{(i)}(\boldsymbol{u})}\right)^2 \nonumber \\
&\leq -2\sum_{\boldsymbol{u} \in \mathcal{T}^{(i)}_T} \sum_{a = 0}^{1}  \dfrac{N_{T}^{(i)}(\boldsymbol{u}, a)}{T} \log\left(\dfrac{P_{\boldsymbol{u}a}(\hat{\boldsymbol{\omega}}^{(i)}_T)}{N_{T}^{(i)}(\boldsymbol{u}, a)/ N_T^{(i)}(\boldsymbol{u})}\right). 
\label{eq:inequality_87}
\end{align}

Employing the Ergodic Theorem once again and combining \eqref{eq:ergodic_theorem_86} and \eqref{eq:inequality_87}, it follows, by assumption \eqref{assumption3:spike_rate_function_bounded}, that, almost surely,
$$\lim_{T \rightarrow \infty} \left|P_{\boldsymbol{u}a}(\hat{\boldsymbol{\omega}}^{(i)}_T) - P_{\boldsymbol{u}a}(\boldsymbol{\omega}_0^{(i)}) \right| = 0.$$

To establish strong consistency of the ML estimator $\hat{\boldsymbol{\omega}}^{(i)}$, we show that $P_{\boldsymbol{u}}: \mathbb{R}^N \rightarrow [0,1]^2$ is injective for all $\boldsymbol{u} \in \mathcal{T}_{T}^{(i)}$ such that $u_{-1}(i) \not= 0$. In this context, for each $\boldsymbol{u} \in \mathcal{T}_{T}^{(i)}$, suppose that two different parameter vectors $\boldsymbol{\omega}^{(i)}$ and $\boldsymbol{\theta}^{(i)}$ lead to the same row vector of transition probabilities, i.e., $P_{\boldsymbol{u}}(\boldsymbol{\omega}^{(i)}) = P_{\boldsymbol{u}}(\boldsymbol{\theta}^{(i)})$. Thus, $P_{\boldsymbol{u}a}(\boldsymbol{\omega}^{(i)}) = P_{\boldsymbol{u}a}(\boldsymbol{\theta}^{(i)})$, for all $a \in \{0,1\}$. 
By assumption \eqref{assumption3:spike_rate_function_bounded}, 
\begin{equation}
P_{\boldsymbol{u}a}(\boldsymbol{\omega}^{(i)}) = P_{\boldsymbol{u}a}(\boldsymbol{\theta}^{(i)}) \Rightarrow \sum_{j=1}^N \left(\omega_{j \rightarrow i} - \theta_{j \rightarrow i}\right)\sum_{s = -\ell + 1}^{-1} \dfrac{u_s(j)}{2^{t - L_t^{(i)} - 1}} = 0.
\label{eq:injective_consequence}
\end{equation}
Without loss of generality, assume that $\omega_{1 \rightarrow i} \not= \theta_{1 \rightarrow i}$ and $\omega_{j \rightarrow i} = \theta_{j \rightarrow i}$ for all $j \in \{2, \ldots, N\}$. In this case, by \eqref{eq:injective_consequence}, we have
$$\left(\omega_{1 \rightarrow i} - \theta_{1 \rightarrow i}\right)\sum_{s = -\ell + 1}^{-1} \dfrac{u_s(j)}{2^{\ell - 1}} = 0,$$
which implies that $\omega_{1 \rightarrow i} = \theta_{1 \rightarrow i}$, since $u_{-1}(i) \not= 0$. This is a contradiction. Therefore, $P_{\boldsymbol{u}}$ is injective.

By assumption  \eqref{assumption3:spike_rate_function_bounded}, the components of $P_{\boldsymbol{u}}$ are continuous. Therefore, since $\boldsymbol{\omega}^{(i)}$ takes values in a compact set, we conclude that, almost surely, 
\begin{equation}
\lim_{T \rightarrow \infty} \left|\hat{\boldsymbol{\omega}}^{(i)}_T - \boldsymbol{\omega}^{(i)}_0\right| = 0,
\label{eq:ML_estimator_as_convergence}
\end{equation}
thus completing the proof. 

\subsubsection{\texorpdfstring{$L^2$}{L2} Consistency}

For each postsynaptic neuron $i \in I$, we know from \eqref{eq:ML_estimator_as_convergence} that, almost surely,
$$\lim_{T \rightarrow \infty} \left|\hat{\boldsymbol{\omega}}^{(i)}_T - \boldsymbol{\omega}^{(i)}_0\right| = 0.$$
Thus, by Assumption \eqref{assumption1:uniform_summability} and the definition \eqref{eq:ML_estimator_definition} of the ML estimator $\hat{\boldsymbol{\omega}}_T^{(i)}$ for the synaptic weight vector $\boldsymbol{\omega}^{(i)}$,  we have that
$$\|\hat{\boldsymbol{\omega}}_T^{(i)} - \boldsymbol{\omega}_0^{(i)}\|^2_2 \leq (\|\hat{\boldsymbol{\omega}}_T^{(i)}\|_2 + \|\boldsymbol{\omega}_0^{(i)}\|_2)^2 < \infty, \hbox{ almost surely},$$
where $\|\cdot\|_2$ denotes the $\ell_2$-norm in $\mathbb{R}^N$.

Therefore, since 
$$\lim_{K \rightarrow \infty} E\left[\|\hat{\boldsymbol{\omega}}_T^{(i)} - \boldsymbol{\omega}_0^{(i)}\|_2^2 \mathbb{I}\left\{\|\hat{\boldsymbol{\omega}}_T^{(i)} - \boldsymbol{\omega}_0^{(i)}\|_2^2 \geq K\right\}\right] = 0,$$
by the dominated convergence theorem, we conclude that
$$\lim_{T \rightarrow \infty} E\left[\|\hat{\boldsymbol{\omega}}_T^{(i)} - \boldsymbol{\omega}_0^{(i)}\|_2^2\right] = 0,$$
thus completing the proof.

\subsection{Proof of Theorem \ref{thm:model_selection_consistency}}

The proof of Theorem \ref{thm:model_selection_consistency} is structured as follows. We first address the overestimation in the proposed model selection process, followed by the treatment of underestimation, and conclude with the proof of consistency.

\subsubsection{Overestimation}

For each postsynaptic neuron $i \in I$, we define 
$$\mathcal{O}_T^{(i)} := \left\{j \in \hat{\mathcal{V}}_T^{(i, \epsilon)} : j \in I - \mathcal{V}^{(i)}\right\}$$
as the event of false positive identification. Using the definition of $\hat{\mathcal{V}}^{(i, \epsilon)}_T$ and applying the union bound, we have that
\begin{equation}
\mathbb{P}\left(\mathcal{O}_T^{(i)}\right) \leq \sum_{j \not\in \mathcal{V}^{(i)}} \mathbb{P}\left[d\left(\hat{\boldsymbol{P}}^{(i)}_{I - \{j\}},  \hat{\boldsymbol{P}}^{(i)}_{I}\right) > \epsilon\right].
\label{eq8.10:false_negative_identification}
\end{equation}

Let us fix $j \not\in \mathcal{V}^{(i)}$. To obtain an upper bound for the right-side of \eqref{eq8.10:false_negative_identification}, we first observe that
\begin{equation}
d\left(\hat{\boldsymbol{P}}^{(i)}_{I - \{j\}},  \hat{\boldsymbol{P}}^{(i)}_{I}\right) > \epsilon \Rightarrow \dfrac{1}{T}\sum_{t=1}^T \left|\hat{P}^{(i)}_t(I - \{j\}) - \hat{P}^{(i)}_t(I)\right| > \epsilon.
\label{eq8.11:distance_definition}
\end{equation}
Since the spike rate function is a Lipschitzian function, there exists a real constant $C > 0$ such that the right side of \eqref{eq8.11:distance_definition} implies that
\begin{equation}
\dfrac{C}{T}\sum_{t=1}^T \sum_{k \in I - \{j\}}\left( \left|\hat{\omega}_{k \rightarrow i}(I - \{j\}) - \hat{\omega}_{k \rightarrow i}(I)\right|\sum_{s = L_t^{(i)} + 1}^{t-1} \frac{x_s(k)}{2^{t - L_t^{(i)} - 1}}\right) > \dfrac{\epsilon}{2}
\label{eq8.12:first_consequence_of_8.11}
\end{equation}
or
\begin{equation}
\dfrac{C}{T}\sum_{t=1}^T \left(\left|\hat{\omega}_{j \rightarrow i}(I)\right|\sum_{s = L_t^{(i)} + 1}^{t-1} \frac{x_s(k)}{2^{t - L_t^{(i)} - 1}}\right) > \dfrac{\epsilon}{2}
\label{eq8.13:second_consequence_of_8.11}
\end{equation}
where, from this point onward, we use the notation 
$$\hat{\boldsymbol{\omega}}^{(i)}_T(F) := (\hat{\omega}_{1 \rightarrow i}(F), \ldots, \hat{\omega}_{N \rightarrow i}(F))$$ 
to denote the ML estimator of $\boldsymbol{\omega}^{(i)}$ obtained by considering only the activity of neurons in the subset $F \subset I$.

For any $k \in I$, we denote the true synaptic weight by $\omega_{k \rightarrow i}^0$, which is an entry of the parameter vector $\boldsymbol{\omega}^{(i)}$. By adding and subtracting $\omega_{k \rightarrow i}^0$ in \eqref{eq8.12:first_consequence_of_8.11}, and applying the triangle inequality, we obtain that
\begin{equation}
\dfrac{C}{T} \sum_{t=1}^T \sum_{k \in I - \{j\}}\left(\left|\hat{\omega}_{k \rightarrow i}(I - \{j\}) - \hat{\omega}_{k \rightarrow i}^0\right|\sum_{s = L_t^{(i)} + 1}^{t-1} \frac{x_s(k)}{2^{t - L_t^{(i)} - 1}}\right) > \dfrac{\epsilon}{4}
\label{eq8.14:first_consequence_of_8.13}
\end{equation}
 or
 \begin{equation}
 \dfrac{C}{T} \sum_{t=1}^T \sum_{k \in I - \{j\}}\left(\left|\hat{\omega}_{k \rightarrow i}(I) - \hat{\omega}_{k \rightarrow i}^0\right|\sum_{s = L_t^{(i)} + 1}^{t-1} \frac{x_s(k)}{2^{t - L_t^{(i)} - 1}}\right) > \dfrac{\epsilon}{4}
 \label{eq8.15:second_consequence_of_8.13}
 \end{equation}

Since $j \not\in V^{(i)}$, we have $\omega_{j \rightarrow i}^0 = 0$, then we can rewrite \eqref{eq8.13:second_consequence_of_8.11} in the following way
\begin{equation}
\dfrac{C}{T}\sum_{t=1}^T \left(\left|\hat{\omega}_{j \rightarrow i}(I) - \omega_{j \rightarrow i}^0\right|\sum_{s = L_t^{(i)} + 1}^{t-1} \frac{x_s(j)}{2^{t - L_t^{(i)} - 1}}\right) > \dfrac{\epsilon}{2}.
\label{eq8.16:first_consequence_of_8.14}
\end{equation}

Using Markov's inequality, after combining \eqref{eq8.14:first_consequence_of_8.13}, \eqref{eq8.15:second_consequence_of_8.13} and \eqref{eq8.16:first_consequence_of_8.14}, we find that, for any $j \not\in V^{(i)}$, $\mathbb{P}\left[d\left(\hat{\boldsymbol{P}}^{(i)}_{I - \{j\}}, \hat{\boldsymbol{P}}_I^{(i)}\right) > \epsilon\right]$ may be bounded above by
\begin{align*}
&\quad\dfrac{4C}{T\epsilon} \sum_{t=1}^T \sum_{k \in I - \{j\}} \left[\left(\sum_{s = L_t^{(i)} + 1}^{t-1} \frac{x_s(k)}{2^{t - L_t^{(i)} + 1}}\right) E\left(\left|\hat{\omega}_{k \rightarrow i}(I - \{j\}) - \hat{\omega}_{k \rightarrow i}^0\right|\right)\right] \\
&+ \dfrac{4C}{T\epsilon} \sum_{t=1}^T \sum_{k \in I - \{j\}} \left[\left(\sum_{s = L_t^{(i)} + 1}^{t-1} \frac{x_s(k)}{2^{t - L_t^{(i)} + 1}}\right) E\left(\left|\hat{\omega}_{k \rightarrow i}(I) - \hat{\omega}_{k \rightarrow i}^0\right|\right)\right] \\
&+ \dfrac{2C}{T\epsilon} \sum_{t=1}^T  \left[\left(\sum_{s = L_t^{(i)} + 1}^{t-1} \frac{x_s(j)}{2^{t - L_t^{(i)} + 1}}\right) E\left(\left|\hat{\omega}_{j \rightarrow i}(I) - \hat{\omega}_{j \rightarrow i}^0\right|\right)\right]. \\
\end{align*}

Therefore, using the aforementioned upper bound, inequality \eqref{eq8.10:false_negative_identification}, $L^2$-consistency of MLE of $\boldsymbol{\omega}_0^{(i)}$ (Theorem \ref{thm:MLE_consistency}), and Ces\'{a}ro's mean, we conclude that
$$\lim_{T \rightarrow \infty} \mathbb{P}\left(\mathcal{O}_T^{(i)}\right) = 0,$$
thereby completing the proof.

\subsubsection{Underestimation}

For each postsynaptic neuron $i \in I$, we define
$$\mathcal{U}_T^{(i)} := \left\{j \not\in \hat{\mathcal{V}}_T^{(i,\epsilon)}: j \in \mathcal{V}^{(i)}\right\}$$
as the event of false negative identification. Using the definition of $\hat{\mathcal{V}}_T^{(i, \epsilon)}$ and applying the union bound, we have that
\begin{equation}
\mathbb{P}\left(\mathcal{U}_T^{(i)}\right) \leq \sum_{j \in \mathcal{V}^{(i)}} \mathbb{P}\left[d\left(\hat{\boldsymbol{P}}^{(i)}_{I - \{j\}}, \hat{\boldsymbol{P}}_I^{(i)}\right) \leq \epsilon\right].
\label{eq8.17:negative_false_identification}
\end{equation}

Let us fix $j \in \mathcal{V}^{(i)}$. To obtain an upper bound for the right-side of \eqref{eq8.17:negative_false_identification}, we first observe that
\begin{equation}
d\left(\hat{\boldsymbol{P}}^{(i)}_{I - \{j\}}, \hat{\boldsymbol{P}}_I^{(i)}\right) \leq \epsilon \Rightarrow \dfrac{1}{T}\sum_{t=1}^T \left|\hat{P}_t^{(i)}(I - \{j\}) - \hat{P}_t^{(i)}(I)\right| \leq \epsilon, 
\end{equation}
which implies that 
\begin{align}
&\dfrac{1}{T} \sum_{t = 1}^T \left|\hat{P}_t^{(i)}(I - \{j\}) - P_t^{(i)}(I - \{j\})\right| - \dfrac{1}{T}\sum_{t=1}^T\left|\hat{P}_t^{(i)}(I) - P_t^{(i)}(I)\right| \nonumber \\
&\quad \geq \dfrac{1}{T}\sum_{t=1}^T \left|P_t^{(i)}(I) - P_t^{(i)}(I-\{j\})\right| - \epsilon.
\label{eq8.19:consequence_8.18}
\end{align}

Define, for each $i \in I$,
$$D_i := \left[\sum_{k \in \mathcal{V}^{(i)}_{-}} \omega_{k \rightarrow i}, \ \sum_{k \in \mathcal{V}^{(i)}_{+}} \omega_{k \rightarrow i}\right],$$
where $\mathcal{V}^{(i)}_{-} := \left\{k \in I: \omega_{k \rightarrow i} < 0\right\}$ and $\mathcal{V}^{(i)}_{+} := \left\{k \in I: \omega_{k \rightarrow i} > 0\right\}.$
Notice that, under the assumptions (1) and (2), this interval is always bounded. Moreover, by assumption (3), we know that the spike rate function is a strictly increasing and uniformly Lipschitz continuous. Then, by the mean value theorem, 
\begin{equation}
\left|P_t^{(i)}(I) - P_t^{(i)}(I-\{j\})\right| \geq \inf_{u \in D_i} \left\{\phi_i^{\prime}(u)\right\}|\omega_{j \rightarrow i}| := m_{ij}.
\label{eq8.20:mean_value_theorem}
\end{equation}
Note that $j \in \mathcal{V}^{(i)}$ implies $\omega_{j \rightarrow i} \not= 0$. Since $\phi_i$ is a strictly increasing function, we have $\inf_{u \in D_i}\{\phi_i^{\prime}(u)\} > 0$. Thus, $m_{ij} > 0$.

By combining \eqref{eq8.19:consequence_8.18} and \eqref{eq8.20:mean_value_theorem}, we obtain that
\begin{equation}
\dfrac{1}{T} \sum_{t = 1}^T \left|\hat{P}_t^{(i)}(I - \{j\}) - P_t^{(i)}(I - \{j\})\right| \geq \dfrac{m_{ij} - \epsilon}{2}.
\label{eq8.21:first_consequence_8.20}
\end{equation}
or
\begin{equation}
 \dfrac{1}{T}\sum_{t=1}^T\left|\hat{P}_t^{(i)}(I) - P_t^{(i)}(I)\right| \geq \dfrac{m_{ij} - \epsilon}{2}.
 \label{8.22:second_consequence_8.20}
 \end{equation}

Using Markov's inequality, after combining \eqref{eq8.21:first_consequence_8.20} and \eqref{8.22:second_consequence_8.20}, we find that, for any $j \in V^{(i)}$, $\mathbb{P}\left[d\left(\hat{\boldsymbol{P}}^{(i)}_{I - \{j\}}, \hat{\boldsymbol{P}}_I^{(i)}\right) \leq \epsilon\right]$ may be bounded above by
\begin{align}
&\dfrac{2}{T(m_{ij} - \epsilon)} \sum_{t=1}^T E\left(\left|\hat{P}_t^{(i)}(I - \{j\}) - {P}_t^{(i)}(I-\{j\})\right|\right) \nonumber \\
&+\dfrac{2}{T(m_{ij} + \epsilon)} \sum_{t=1}^T E\left(\left|\hat{P}_t^{(i)}(I - \{j\}) - {P}_t^{(i)}(I)\right|\right).
\label{eq8.23:upper_bound_P}
\end{align}
Since the spike rate function is a Lipschitzian function, there exists a real constant $C > 0$ such that \eqref{eq8.23:upper_bound_P} can be bounded above by
\begin{align}
&\dfrac{2C}{T(m_{ij} + \epsilon)} \sum_{t=1}^T\sum_{k \in I-\{j\}} E\left[\left|\hat{\omega}_{k \rightarrow i}(I - \{j\}) - \omega_{k \rightarrow i}^0\right|\right] \nonumber \\
&+ \dfrac{2C}{T(m_{ij} + \epsilon)} \sum_{t=1}^T\sum_{k \in I} E\left[\left|\hat{\omega}_{k \rightarrow i}(I) - \omega_{k \rightarrow i}^0\right|\right].
\label{eq8.24:upper_bound_8.23}
\end{align}

Therefore, using the aforementioned upper bound, inequality \eqref{eq8.17:negative_false_identification}, $L^2$-consistency of MLE for $\boldsymbol{\omega}_0^{(i)}$ (Theorem \ref{thm:MLE_consistency}), and Ces\'{a}ro's mean, we conclude that
$$\lim_{T \rightarrow \infty} \mathbb{P}\left(\mathcal{U}_T^{(i)}\right) = 0,$$
thereby completing the proof.

\subsubsection{Consistency}

We observe that
$$\left\{\hat{\mathcal{V}}^{(i, \epsilon)}_T \not= \mathcal{V}^{(i)}\right\} = \mathcal{O}_T^{(i)} \cup \mathcal{U}_T^{(i)}.$$
Thus, 
$$0 \leq \lim_{T \rightarrow \infty} \mathbb{P}\left(\hat{\mathcal{V}}_T^{(i)} \not= \mathcal{V}_T^{(i)}\right) \leq \lim_{T \rightarrow \infty} \mathbb{P}\left(\mathcal{O}_T^{(i)}\right) + \lim_{T \rightarrow \infty} \mathbb{P}\left(\mathcal{U}_T^{(i)}\right) = 0,$$
which follows from overestimation and underestimation results, thereby completing the proof.

\section{Final remarks}
\label{sec10}

{The brain is one of the most complex systems ever studied, with approximately 86 billion neurons and trillions of synapses \citep{herculano2009human}. Only recently have recording methods advanced enough to access multi-unit and multi-variable neural data, such as with multi-electrode arrays \citep{thomas1972miniature, morin2005investigating} and voltage-imaging techniques \citep{peterka2011imaging}, alongside optogenetic approaches for precise stimulation \citep{fenno2011development}. However, even in this era of big data, much remains unknown about how this vast array of recorded signals interacts to generate behavior. In this paper, we contribute to the understanding of neuronal connectivity by leveraging stochastic models to capture the probabilistic nature of spike interactions within neural circuits. The model we employed, which incorporates variable-length memory, aligns well with known biological phenomena.} 

\textcolor[rgb]{0,0,0}{
Recent advances in spiking neural networks (SNNs) have led to increasingly sophisticated models and training strategies that capture complex spatiotemporal neuronal interactions. For instance, directly trained deep SNNs \citep{zheng2021going} and high-order information bottleneck approaches \citep{yang2023effective} demonstrate the feasibility of scaling up spike-based learning while retaining efficiency. Meanwhile, techniques such as deep residual learning \citep{fang2021deep} and self-supervised methods \citep{yang2024self} highlight how architectural innovations can stabilize training and enhance performance on challenging tasks, including robust event-based optical flow estimation. Moreover, nonlinear and maximum entropy formulations \citep{yang2023snib, yang2024maximum} underscore emerging theoretical perspectives on the role of information bottlenecks in spike-based computation. In this context, our approach to identifying functional interactions among neurons with variable-length memory complements these developments by offering a statistically grounded method to reveal and quantify the pairwise synaptic influences tested in high-dimensional, high-resolution neural data. Such consistent model selection provides a principled framework to validate or refine the connectivity assumptions underlying advanced SNN architectures, offering deeper insight into how neuronal interactions shape the dynamics of spiking systems.}

{From the perspective of stochastic sources, \textcolor[rgb]{0,0,1}{our} model captures a broad spectrum of brain stochasticity, accounting for three primary sources: channel noise, synaptic noise, and network noise \citep{faisal2008noise}. This inherent randomness can be challenging when dealing with deterministic dynamics, although stochastic differential equation models, such as stochastic leaky integrate-and-fire neuron models, have been successful in various applications \citep{lansky2008review, sacerdote2013stochastic}. However, the leaky term in these models introduces a strong correlation effect that is not observed in our case due to the variable length-memory.} \textcolor[rgb]{0,0,0}{By successfully modeling stochastic neurons with variable-length memory, our results lend support to the hypothesis that network-level neuronal interactions cannot be fully explained by classical Markovian approaches \citep{truccolo2005point,galves2013infinite}. This is significant because it supports the notion that neural coding strategies may depend substantially on memory-dependent processes, challenging simpler models of neuron dynamics. }

{To address the intricacies of network connectivity, we explored factors such as sparseness and network size. By dividing our study into specific scenarios, we examined how these characteristics influence estimation accuracy, noting that both can have positive or negative impacts. Additionally, the balance of excitation and inhibition led to greater discrepancies in estimation. This challenge is similarly observed in experimental recordings, where factors like connection strength, recording duration, and the degree of sparseness within the brain region often need more consideration. Our scenarios suggest ways to address these challenges and improve estimation accuracy.}

\textcolor[rgb]{0,0,0}{Despite demonstrating consistent and robust performance, the proposed method is subject to some limitations. One of them is that our approach relies on discrete-time binning, and the choice of bin size may significantly influence estimation accuracy. Furthermore, although the model effectively captures variable-length memory interactions, it does not explicitly incorporate synaptic plasticity or other adaptive changes in network connectivity over longer time scales.}

\textcolor[rgb]{0,0,0}{In terms of computational complexity, the primary cost of the proposed method arises from calculating the sensitivity measure $d$ for each pair of neurons, which involves computing empirical spiking probabilities. Given a network with $N$ neurons, identifying pairwise interactions results in an overall complexity that scales quadratically with the number of neurons. This is because, for each neuron, we evaluate its interactions with all other neurons in the network. Such quadratic dependence on $N$ is a common characteristic of network inference methods. While this scaling remains manageable for small to medium-sized networks, a shortcoming is that it may become computationally intensive for very large networks. Quantitatively speaking, our simulations suggest that networks with up to a few hundred neurons are still feasible, given moderate-length recordings (seconds). However, for larger networks, the increased computational cost and memory requirements may become limiting factors. The method's performance also depends on the density of the network. Sparse networks, which are biologically plausible in the cortex for example, are computationally less demanding, while denser networks may require additional approximations to maintain efficiency.} 

\textcolor[rgb]{0,0,0}{To mitigate these limitations, potential optimizations include parallel processing and efficient memory management, which would allow the method to handle larger networks on standard hardware. For even larger networks, we recommend using distributed computing frameworks or applying dimensionality reduction techniques to reduce the number of pairwise comparisons. These approaches represent promising directions for future work.}

\textcolor[rgb]{0,0,0}{Although the maximum likelihood estimator has been the main focus of this work, other estimation methods, such as moment-based methods, Bayesian estimators, or machine learning techniques, could have been considered. However, we chose the maximum likelihood estimator primarily due to the possibility of mathematically demonstrating its consistency for the neuronal model adopted in this research (see Theorem \ref{thm:MLE_consistency}) We believe that the theoretical support, combined with the results obtained in the simulations, reinforces the robustness and reliability of the method. Similarly, an evaluation of the proposed model selection method in comparison to established approaches in the literature, such as information criteria \citep{de2022estimating} or regularization methods \citep{ost2020sparse}, could provide additional insights into its advantages and limitations. However, given the complexity and scope required for a fair and detailed comparison, we believe this topic deserves a dedicated study. This study represents a promising direction for future work, enabling a deeper understanding of specific scenarios in which each approach stands out.}
  
{For our data application, we chose to work with neurons from the CA1 region of the hippocampus, a brain area critical for memory, learning, and spatial navigation \citep{Buz02}. Establishing a relationship between behavior, neural firing (supra-threshold activity), and membrane electrical signals (sub-threshold activity) has long been hindered by the technical challenges of simultaneously analyzing different types of brain activity, even with commonly used methods like calcium imaging and multi-electrode recording. Our collaborative research, which integrates advanced theoretical and statistical methods for estimating functional interactions among stochastic neurons, openly available extracellular recordings from rats, and biophysical computational modeling, has successfully generated spiking patterns that resemble those of CA1 neurons. This offers a novel perspective on how to approach these functional roles in neurons.
}

{While our data choice focused on single-neuron signals, the methodology for estimating functional interactions could be extended to capture higher levels of organization, such as electroencephalogram (EEG) signals or local field potentials (LFPs). In these cases, the modeling approach could be adapted to firing-rate models, facilitating broader applications in studying neural dynamics at the population level. Here, connections would represent population-level connectivity, similar to how anatomical maps reflect the density of white matter. Our stochastic models are well-suited to accommodate higher noise sources, including external and environmental noise. Future research should explore these possibilities to deepen our understanding of functional interactions.}

\section*{Acknowledgments} 
R.F.F. was partially supported by grant $\#2018/25076-3$, S\~{a}o Paulo Research Foundation (FAPESP). R.F.O.P. was supported by the Palm Health-Sponsored Program in Computational Brain Science and Health, FAU Stiles-Nicholson Brain Institute, and the Jupiter Life Science Initiative (JLSI). This work is also part of the activities of the FAPESP Research, Innovation and Dissemination Center for Neuromathematics ($\#2013/07699-0$).

\bibliographystyle{elsarticle-harv} 
\bibliography{bibliography.bib}

\end{document}